\documentclass[12pt]{article}

\usepackage[totalwidth=460truept,totalheight=608truept]{geometry}
\usepackage{latexsym,graphicx,amsfonts,amssymb,amsmath}
\usepackage[hypertexnames=false,hidelinks]{hyperref}

\def\theequation{\arabic{section}.\arabic{equation}}

\renewcommand{\theequation}{\thesection.\arabic{equation}}
\global\arraycolsep=1truept
\renewcommand{\theequation}{\arabic{section}.\arabic{equation}}

\begin{document}

\null

\vskip.1truecm

\begin{center}
{\huge \textbf{Perturbation Spectra and}}

\vskip.75truecm

{\huge \textbf{Renormalization-Group Techniques}}

\vskip.75truecm

{\huge \textbf{in Double-Field Inflation and}}

\vskip1truecm

{\huge \textbf{Quantum Gravity Cosmology}}

\vskip1truecm

\textsl{Damiano Anselmi}

\vskip .1truecm

\textit{Dipartimento di Fisica \textquotedblleft Enrico Fermi", Universit%
\`{a} di Pisa}

\textit{Largo B. Pontecorvo 3, 56127 Pisa, Italy,}

\textit{INFN, Sezione di Pisa,}

\textit{Largo B. Pontecorvo 3, 56127 Pisa, Italy,}

\textit{Laboratory of High Energy and Computational Physics, NICPB, }

\textit{R\"{a}vala 10, Tallinn 10143, Estonia}

damiano.anselmi@unipi.it

\vskip.5truecm

\textbf{Abstract}
\end{center}

We study primordial cosmology with two scalar fields that participate in
inflation at the same time, by coupling quantum gravity (i.e., the theory $%
R+R^{2}+C^{2}$ with the fakeon prescription/projection for $C^{2}$) to a
scalar field with a quadratic potential. We show that there exists a
perturbative regime that can be described by an asymptotically de Sitter,
cosmic RG flow in two couplings. Since the two scalar degrees of freedom mix
in nontrivial ways, the adiabatic and isocurvature perturbations are not RG
invariant on superhorizon scales. It is possible to identify the correct
perturbations by using RG invariance as a guiding principle. We work out the
resulting power spectra of the tensor and scalar perturbations to the NNLL
and NLL orders, respectively. An unexpected consequence of RG invariance is
that the theory remains predictive. Indeed, the scalar mixing affects only
the subleading corrections, so the predictions of quantum gravity with
single-field inflation are confirmed to the leading order.

\vfill\eject

\section{Introduction}

\label{intro}\setcounter{equation}{0}

Inflation explains the approximate isotropy and homogeneity of the cosmic
microwave background radiation by means of a primordial accelerated
expansion of the universe \cite%
{englert,starobinsky,kazanas,sato,guth,linde,steinhardt,linde2}, which
originates the present large-scale structure from the quantum fluctuations 
\cite{mukh,mukh2,hawk,guth2,staro2,bardeen,mukh3}. In the most popular
approach, the expansion is driven by a scalar field rolling down a potential 
\cite{weinberg,reviews,defelice}, which in many cases leads to a scalar
perturbation spectrum that agrees with observations \cite{encicl,Planck18}.
There also exists a \textquotedblleft geometric\textquotedblright\ approach,
where inflation is driven by the metric itself, as in the Starobinsky $%
R+R^{2}$ model \cite{starobinsky} and the $f(R)$ theories \cite%
{defelice,otherfR}. In a third\ approach, instead \cite%
{CMBrunning,FakeOnScalar}, inflation is described as a cosmic
renormalization-group (RG) flow, triggered by the dependence of the
background metric (rather than the radiative corrections). The main types of
flows can be classified according to the properties of their spectra,
without referring to their origins from specific actions or models \cite{AFP}%
.

The $R+R^{2}$ theory works well phenomenologically. However, once we include 
$R^{2}$ there is no compelling argument for excluding the square $C_{\mu \nu
\rho \sigma }C^{\mu \nu \rho \sigma }\equiv C^{2}$ of the Weyl tensor $%
C_{\mu \nu \rho \sigma }$, which has the same dimension in units of mass.
Once both are present, we do not need to add further terms, since the
resulting classical action%
\begin{equation}
S_{\text{QG}}=-\frac{1}{16\pi G}\int \mathrm{d}^{4}x\sqrt{-g}\left( R+\frac{1%
}{2m_{\chi }^{2}}C_{\mu \nu \rho \sigma }C^{\mu \nu \rho \sigma }-\frac{R^{2}%
}{6m_{\phi }^{2}}\right)  \label{sqg}
\end{equation}
is renormalizable by power counting \cite{stelle}. However, $C^{2}$ causes
difficulties, if treated as usual, because it gives ghosts, which are
responsible for unacceptable physics. Even if the ghosts are assumed to be
heavy or short-lived, it is always possible to imagine situations where such
assumptions are ineffective and the internal consistency of the theory
breaks down. An example is primordial cosmology, where we must properly
treat the high-energy limit (through the Bunch-Davies vacuum condition, for
example), even if our purpose is just to make predictions about the
low-energy limit (which means the superhorizon limit, in this case).

The problem of ghosts is due to an incorrect quantization procedure and
disappears \cite{LWgrav} once we quantize the associated degrees of freedom
as purely virtual particles, or fakeons \cite{fakeons}. The physical
predictions change, because the new prescription changes the very definition
of Feynman diagram. It does so in the only way that is consistent with the
optical theorem and unitarity, hence the theory obtained from (\ref{sqg}) is
both renormalizable and unitary. In particular, the renormalization is
unaffected by the prescription\footnote{%
The one-loop beta functions can be found in \cite{betaHD,UVQG}.}.

In the end, the theory contains a unique purely virtual particle, with spin
2 and mass $m_{\chi }$. We denote it by $\chi _{\mu \nu }$. It is removed
from the physical spectrum at all energies by \textquotedblleft integrating
it out\textquotedblright\ with the fakeon prescription. Similar operations
provide a new formulation of Lee-Wick theories \cite{PivaLW} and overcome
the difficulties of their original formulation.

Cosmology is an arena where we can appreciate how important these operations
are in order to get to sensible physics, because, as said, it is not
possible to make predictions about the low-energy (superhorizon) limit,\ if
we do not properly take care of the opposite limit. This requires a theory
that is consistent (as a perturbative quantum field theory) at all energies.

We can neglect the loop corrections \cite{weinberg}, but other challenges
appear. For example, the quantization must be performed on a nontrivial
background. This can be done by first classicizing quantum gravity \cite%
{classicization,FLRW} and then quantizing the (projected) classical limit by
means of the Bunch-Davies quantization condition. The consistency of this
procedure on the FLRW background leads to the bound $m_{\chi }>m_{\phi }/4$
on the coefficients of the terms $R^{2}$ and $C^{2}$ \cite{ABP}, which makes
the prediction on the tensor-to-scalar ratio $r$ quite sharp ($0.4\lesssim
1000r\lesssim 3$ with 60 efoldings), at least in the case of single-field
inflation.

Although the most popular models of inflation involve a single scalar field,
the possibility that more scalars participate in inflation at the same time
is not excluded and has attracted considerable attention in the literature 
\cite{multifinfl,gundhi,adiabiso}. Double-field and multi-field inflation
are also studied because of their interesting theoretical aspects. Due to
their complexity, they are normally handled by means of numerical methods.

In this paper, we minimally couple the quantum gravity theory defined as
explained above to a matter scalar field $\varphi $ with a quadratic
potential. The theory so obtained is renormalizable \cite{Absograv}. We
think that this makes it worth of consideration, given that
renormalizability is the criterion provided by high-energy physics to
overcome the arbitrariness of classical theories. We study a scenario where
both scalar fields ($\varphi $ and the one due to $R^{2}$, which we denote
by $\phi $) participate in inflation at the same time and identify a
perturbative regime where we can apply the approach of refs. \cite%
{CMBrunning,FakeOnScalar}, based on the idea of cosmic RG flow. We recall
that\ this type of flow, originated by the dependence on the background
metric, has an asymptotically de Sitter beta function and is such that the
spectra of the cosmological perturbations satisfy equations of the
Callan-Symanzik type in the superhorizon limit.

We calculate the spectrum of the tensor fluctuations to the
next-to-next-to-leading log (NNLL) order, which is given in formula (\ref{pT}%
), and the spectra of the scalar fluctuations to the NLL order, given in
formula (\ref{calP}). As usual, the spectra of the tensor perturbations are
affected by $C^{2}$ from the zeroth order, as emphasized by the dependence
of the results on $m_{\chi }$. The spectra of the scalar perturbations are
not affected by $m_{\chi }$ to the NLL order included, where they coincide
with those of the Starobinsky theory, which is obtained by taking the limit $%
m_{\chi }\rightarrow \infty $.

Along the way, we meet new challenges. The first difficulty is to identify
the right couplings. The quantity $\epsilon =-\dot{H}/H^{2}$ helps us
identify one of them, which we denote by $\alpha $, but the identification
of the second one, which we denote by $\lambda $, is not straightforward.
The answer is provided by the perturbation spectra, if we require that

$a$) the beta functions are power series in $\alpha $ and $\lambda $ and
start quadratically;

$b$) the flow is asymptotically de Sitter in the infinite past, which means
that $\alpha $ and $\lambda $ tend to zero for $t\rightarrow -\infty $,
where the metric tends to the de Sitter one;

$c$) the perturbation spectra are power series in $\alpha $ and $\lambda $,
apart from overall factors;

$d$) the perturbation spectra are RG\ invariant on superhorizon scales.

Requirement $a$) means that the beta functions behave like those of an
asymptotically free quantum field theory, for example QCD. The possibility
of having overall factors with negative powers (or even fractional powers
and essential singularities \cite{AFP}) is due to the fact that the spectra
of the scalar perturbations commonly have such a feature.

A second issue is that in double-field inflation the two scalar modes, $\phi 
$ and $\varphi $, mix in nontrivial ways. This makes the identification of
the right physical quantities challenging. Specifically, the adiabatic and
isocurvature perturbations $\mathcal{R}_{\text{adiab}}$ and $\mathcal{R}_{%
\text{iso}}$, as commonly defined \cite{gundhi,adiabiso}, are not RG\
invariant, which means that they are not conserved on superhorizon scales.
Nevertheless, we can use RG invariance to identify conserved combinations $%
\mathcal{R}_{\text{RG}}^{(1)}$ and $\mathcal{R}_{\text{RG}}^{(2)}$. The
right couplings $\alpha $ and $\lambda $, the beta functions, and the
spectra of $\mathcal{R}_{\text{RG}}^{(1)}$ and $\mathcal{R}_{\text{RG}}^{(2)}
$ must be identified altogether, in order to satisfy the requirements $a$-$d$%
). Once this goal is achieved, the RG properties are exploited to the
fullest. The lesson is that cosmic RG invariance is a physical principle to
be placed side by side with gauge invariance.

The third issue concerns the predictivity of double-field inflation. Before
explaining this point, it is worth to briefly comment on the predictivity of
single-field inflation in quantum gravity with purely virtual particles. We
have already remarked that, although the typical predictions of primordial
cosmology concern the superhorizon limit, it is not enough to work around
that limit to make such predictions. This is in contrast to what happens in
high-energy physics, where a low-energy effective theory is enough to make
predictions about low energies.

There are two main reasons why we must include the high energies to make
predictions about the low energies in primordial cosmology. The first reason
is that on a nontrivial background we need to provide a quantization
condition. This goal is normally achieved by means of the Bunch-Davies
vacuum condition, which refers to the high-energy limit of the theory, where
the problem can be handled because it reduces to a flat-space one in
conformal time. Clearly, it makes sense to impose such a condition only if
the theory is consistent at high energies, which is not true when it
contains ghosts, because they do not disappear there. Fakeons, instead,
disappear everywhere, which leads us to the second reason why it is crucial
to include the high energies to make predictions about the low energies. In
the low energy regime fakeons disappear for free, because they massive, but
in the opposite limit we must impose a condition to ensure that they
disappear as well. This condition, which follows from the consistency of the
fakeon projection and in particular its classicization \cite%
{classicization,FLRW} on a curved background, is the bound $m_{\chi
}>m_{\phi }/4$ found in ref. \cite{ABP} (to which we refer as
\textquotedblleft ABP\ bound\textquotedblright ). A bonus of the ABP bound
is that it makes the physical predictions quite sharp in single-field
inflation. Note that in flat space there would not be such issues.

In double-field inflation the ABP\ bound $m_{\chi }>m_{\phi }/4$ remains the
same, but in principle the dependence on the second coupling could spoil the
relation between the tensor-to-scalar ratio $r$ and the scalar tilt $n_{%
\text{s}}$. An unexpected result of RG invariance is that, instead, this
does not happen, at least to the leading order (in the perturbative regime
that we identify in the next section), where the final prediction agrees
with the one of the pure quantum gravity theory. This crucial property is
satisfied by the RG invariant scalar perturbations $\mathcal{R}_{\text{RG}%
}^{(1)}$ and $\mathcal{R}_{\text{RG}}^{(2)}$, not by the standard adiabatic
and isocurvature perturbations $\mathcal{R}_{\text{adiab}}$ and $\mathcal{R}%
_{\text{iso}}$. In the end, the predictions of the pure theory found in \cite%
{ABP} turn out to be quite robust.

Other troubles with double- and multi-field inflation are known. For
example, the curvature perturbations $\mathcal{R}_{\text{adiab}}$ and $%
\mathcal{R}_{\text{iso}}$ are not conserved in between the horizon exit and
the horizon re-entry. This problem must be reconsidered now, given that $%
\mathcal{R}_{\text{adiab}}$ and $\mathcal{R}_{\text{iso}}$ are not RG
invariant and the right physical quantities are $\mathcal{R}_{\text{RG}%
}^{(1)}$ and $\mathcal{R}_{\text{RG}}^{(2)}$.

We work in the \textquotedblleft geometric framework\textquotedblright\
introduced in ref. \cite{ABP}, which is the one where the higher-derivative
terms $R^{2}$ and $C^{2}$ of the action are kept as such. Apart from the
unusual presence of the term $C^{2}$, the geometric framework coincides with
the familiar Jordan frame. It allows us to treat double-field inflation with
just one explicit scalar field.

Alternatively, we could use the \textquotedblleft inflaton
framework\textquotedblright , which is obtained by introducing the inflaton
field $\phi $ explicitly to eliminate $R^{2}$. It coincides with the
Einstein frame, apart from the presence of $C^{2}$. We prefer not to use
this approach here, since it complicates the two-scalar potential $\mathcal{V%
}(\phi ,\varphi )$, by mixing a potential of class I with a potential of
class II,\ according to the classification recently introduced in ref. \cite%
{AFP}.

A third framework is available, where the scalar $\phi $ and the spin-2
fakeon $\chi _{\mu \nu }$ are introduced explicitly to eliminate both
higher-derivative terms $R^{2}$ and $C^{2}$ \cite{Absograv}. At the
practical level, it is not very convenient for single-field inflation \cite%
{ABP} and, more generally, cosmology \cite{FLRW}.

The interest in a theory like the one we study here is that it is
renormalizable, besides being local and unitary, so it can help address the
issue of predictivity in quantum gravity when matter is present. As said,
the outcome is that the predictions of the pure theory are not affected to
the leading order. This is a nontrivial consequence of RG invariance.
Simpler models of double-field inflation can be considered, if we relax the
renormalizability constraint.

We mention that earlier calculations of the running of spectral indices in
various scenarios (without fakeons) can be found in \cite{run1} and
calculations of subleading corrections are available in \cite{run2} and \cite%
{run3} in single-field and multi-field inflation, respectively.

The paper is organized as follows. In section \ref{betaf} we introduce the
theory, derive the beta functions of the cosmic RG flow and recall the RG
properties of the spectra. In section \ref{pertu}, we present the action and
the fluctuations. In section \ref{tensorQG} we compute the spectrum of the
tensor perturbations to the NNLL\ order. In section \ref{scalarocm} we study
the spectra of the scalar perturbations to the NLL order, working out the RG
invariant combinations $\mathcal{R}_{\text{RG}}^{(1)}$ and $\mathcal{R}_{%
\text{RG}}^{(2)}$. In section \ref{predictions} we summarize the results and
comment on the physical predictions of the model. Section \ref{conclusions}
contains the conclusions, while the appendices collect various reference
formulas that are too involved to be displayed in the main sections of the
paper.

\section{Cosmic RG\ flow for double-field inflation}

\setcounter{equation}{0}\label{betaf}

In this section we formulate the cosmic RG flow in two couplings, associated
with the model of double-field inflation that we study in the paper. We
recall that the \textquotedblleft cosmic\textquotedblright\ RG flow is not
due to the radiative corrections. It is an alternative approach to inflation
itself. In particular, the role of the sliding scale is played by minus the
conformal time $\tau $, so RG invariance and conservation are mapped into
each other. This is the reason why RG invariance allows us to find the
conserved quantities on superhorizon scales. The inflaton framework and the geometric
framework mentioned above (which correspond to the usual Einstein frame and
Jordan frame, respectively, apart from the presence of the Weyl-squared
term) are viewed as different schemes for the flow. Like in quantum field
theory, RG invariance ensures that the physical quantities (such as the
spectra) are scheme independent. The main types of cosmic RG flows in
single-field inflation have been classified in \cite{AFP}.

The action of the model is%
\begin{eqnarray}
S_{\text{QG+scal}} &=&-\frac{1}{16\pi G}\int \mathrm{d}^{4}x\sqrt{-g}\left(
R+\frac{1}{2m_{\chi }^{2}}C_{\mu \nu \rho \sigma }C^{\mu \nu \rho \sigma }-%
\frac{R^{2}}{6m_{\phi }^{2}}\right)  \notag \\
&&+\int \mathrm{d}^{4}x\sqrt{-g}\left( \frac{1}{2}D_{\mu }\varphi D^{\mu
}\varphi -\mathcal{V}(\varphi )+Q(\varphi )R\right) .  \label{Sqgscal}
\end{eqnarray}%
where $\varphi $ is the extra scalar field. Assuming for simplicity
invariance under the symmetry $\varphi \rightarrow -\varphi $,
renormalizability requires%
\begin{equation}
\mathcal{V}(\varphi )=\frac{m^{2}}{2}\varphi ^{2}+\frac{f_{4}}{4!}\varphi
^{4},\qquad Q(\varphi )=f_{2}\varphi ^{2},  \label{reno}
\end{equation}%
where $m$, $f_{2}$ and $f_{4}$ are constants. Throughout the paper, we
assume $f_{2}=f_{4}=0$, which leaves the theory renormalizable\footnote{%
Renormalization does not turn on $f_{2}$ and $f_{4}$ when they are both
absent at the classical level \cite{Absograv}. The reason is that at $%
f_{2}=f_{4}=0$ each external $\varphi $ leg carries a partial derivative $%
\partial _{\mu }$ (vertices due to $\sqrt{-g}g^{\mu \nu }\partial _{\mu
}\varphi \partial _{\nu }\varphi $) or a mass $m$ (vertices due to $%
m^{2}\varphi ^{2}\sqrt{-g}$). Thanks to this, the superficial degrees of
divergence of the Feynman diagrams are lowered enough to forbid the
counterterms $R\varphi ^{2}$ and $\varphi ^{4}$. Instead, when either $f_{2}$
or $f_{4}$ is present at the classical level, the other one is generated at
the quantum level \cite{freedman}.}. The choice $f_{2}=f_{4}=0$ allows us to
reduce the computational effort, which is already considerable, without
losing the main properties of double-field inflation.

We look for a background solution where the metric is $g_{\mu \nu }=$diag$%
(1,-a^{2},-a^{2},-a^{2})$ and $\varphi $ depends only on time, by isotropy
and homogeneity. The Friedmann equations and the $\varphi $ equation are%
\begin{equation}
H^{2}+\frac{1}{m_{\phi }^{2}}\left( 6H^{2}\dot{H}-\dot{H}^{2}+2H\ddot{H}%
\right) =\frac{\hat{\kappa}^{2}}{4}\left( \dot{\varphi}^{2}+2\mathcal{V}%
\right) ,\qquad \ddot{\varphi}+3H\dot{\varphi}+\mathcal{V}^{\prime }=0,
\label{fried}
\end{equation}%
where $a(t)$ denotes the scale factor, $H=\dot{a}/a$ is the Hubble parameter
and $\hat{\kappa}\equiv \sqrt{16\pi G/3}$. We omit the other Friedmann
equation, since it can be obtained (multiplied by $H$) from the equations
just written, by differentiating the first one with respect to time and
using the second one to eliminate $\ddot{\varphi}$.

Before proceeding, we recall that, although $H$ is almost constant during
inflation, its time dependence is crucial for the spectra of the
fluctuations, which are normally expressed by means of the quantity $%
\epsilon =-\dot{H}/H^{2}$ (which here becomes one of the couplings). Higher
time derivatives of $H$, on the other hand, parametrize the higher-order
corrections to the spectra and can be converted into higher powers of the
couplings.

We introduce the couplings $\alpha $, $\varsigma $ and $\lambda $, defined
by 
\begin{equation}
\alpha =-\frac{\dot{H}}{H^{2}},\qquad \varsigma =\frac{\hat{\kappa}\dot{%
\varphi}}{2H},\qquad \lambda =\frac{\varsigma ^{2}}{\alpha }=-\frac{\hat{%
\kappa}^{2}}{4}\frac{\dot{\varphi}^{2}}{\dot{H}},  \label{alfe}
\end{equation}%
together with their beta functions 
\begin{equation}
\beta _{\alpha }=\frac{\mathrm{d}\alpha }{\mathrm{d}\ln |\tau |}=-\frac{\dot{%
\alpha}}{vH},\qquad \beta _{\varsigma }=\frac{\mathrm{d}\varsigma }{\mathrm{d%
}\ln |\tau |}=-\frac{\dot{\varsigma}}{vH},\qquad \beta _{\lambda }=\frac{%
\mathrm{d}\lambda }{\mathrm{d}\ln |\tau |}=-\frac{\dot{\lambda}}{vH}=\frac{%
2\varsigma \beta _{\varsigma }}{\alpha }-\frac{\lambda \beta _{\alpha }}{%
\alpha },  \label{bete}
\end{equation}%
where $v\equiv -(aH\tau )^{-1}$ and $\tau $ denotes the conformal time%
\begin{equation}
\tau =-\int_{t}^{+\infty }\frac{\mathrm{d}t^{\prime }}{a(t^{\prime })}.
\label{tau}
\end{equation}

We first work in the variables $\alpha $ and $\varsigma $, which appear to
be the natural ones at first. We find that $\beta _{\alpha }$ and $\beta
_{\varsigma }$ are power series in $\alpha $ and $\varsigma $, but the other
functions involved in the calculations, such as $H$, $\varphi $ and the
perturbation spectra, are singular for $\alpha \rightarrow 0$, and not just
by overall factors. Yet, it turns out that whenever negative powers of $%
\alpha $ appear, they are associated with positive powers of $\varsigma $,
which suggests that some combination of $\varsigma $ and $\alpha $ is the
correct variable, instead of $\varsigma $.

It turns out that the correct variables are $\alpha $ and $\lambda $. Apart
from some appearances of $\sqrt{\lambda }$ in overall factors, the action of
the tensor fluctuations and the Mukhanov-Sasaki (MS) action of the scalar
fluctuations depends on $\alpha $ and $\lambda $ perturbatively (both in the
limit of infinitely heavy fakeon and at finite $m_{\chi }$). Second, the
beta functions $\beta _{\alpha }(\alpha ,\lambda )$ and $\beta _{\lambda
}(\alpha ,\lambda )$ are regular and have the forms%
\begin{eqnarray}
\beta _{\alpha }(\alpha ,\lambda ) &=&\alpha ^{2}\times (\text{power series
in }\alpha \text{ and }\lambda )=-2\alpha ^{2}+\text{higher orders},  \notag
\\
\beta _{\lambda }(\alpha ,\lambda ) &=&\alpha \lambda \times (\text{power
series in }\alpha \text{ and }\lambda )=-2\alpha \lambda (1-2\varrho )+\text{%
higher orders},  \label{baw}
\end{eqnarray}%
where $\varrho =m^{2}/m_{\phi }^{2}$. Third, there is a region in parameter
space, which is $\varrho \,<1/2$, i.e.,%
\begin{equation}
m<\frac{m_{\phi }}{\sqrt{2}},  \label{bond}
\end{equation}%
where the fixed point $\alpha =\lambda =0$, which is de Sitter space, is
asymptotically free. The running couplings tend to it as%
\begin{equation}
\alpha (-\tau )\simeq \frac{\bar{\alpha}}{1+2\bar{\alpha}\ln |\tau |},\qquad
\lambda (-\tau )\simeq \bar{\lambda}\left( \frac{\alpha (-\tau )}{\bar{\alpha%
}}\right) ^{1-2\varrho },  \label{llrc}
\end{equation}%
for $\tau \rightarrow -\infty $ (which means $t\rightarrow -\infty $), where 
$\bar{\alpha}$ and $\bar{\lambda}$ are integration constants. Finally, the
perturbation spectra are power series in $\alpha $ and $\lambda $, apart
from overall factors. In the rest of the paper we prove these statements and
work out the RG invariant perturbation spectra.

As said, we start by viewing $v$, $H$, $\varphi $ and $\mathcal{V}$ as
functions of $\alpha $ and $\varsigma $. Denoting their partial derivatives
with respect to the couplings by means of subscripts ($v_{\alpha }=\partial
v/\partial \alpha $, $v_{\varsigma }=\partial v/\partial \varsigma $, etc.)
and converting the time derivatives by means of the identity 
\begin{equation*}
\frac{\mathrm{d}}{\mathrm{d}t}=\frac{\mathrm{d}}{a\mathrm{d}\tau }=-vH\frac{%
\mathrm{d}}{\mathrm{d}\ln |\tau |}=-vH\left( \beta _{\alpha }\frac{\partial 
}{\partial \alpha }+\beta _{\varsigma }\frac{\partial }{\partial \varsigma }%
\right) ,
\end{equation*}%
we find the equations%
\begin{equation}
v_{\alpha }\beta _{\alpha }+v_{\varsigma }\beta _{\varsigma }=1-v-\alpha
,\qquad \frac{H_{\alpha }}{H}\beta _{\alpha }+\frac{H_{\varsigma }}{H}\beta
_{\varsigma }=\frac{\alpha }{v},\qquad \varphi _{\alpha }\beta _{\alpha
}+\varphi _{\varsigma }\beta _{\varsigma }=-\frac{2\varsigma }{\hat{\kappa}v}%
.  \label{vhf}
\end{equation}%
The first equation is obtained by differentiating the very definition of $%
v=-(aH\tau )^{-1}$, while the second and third equations follow from the
definitions (\ref{alfe}). Using (\ref{fried}), the potential $\mathcal{V}$
and its derivative $\mathcal{V}^{\prime }$ with respect to $\varphi $ are
given by%
\begin{equation}
\mathcal{V}=\frac{4H^{4}}{\hat{\kappa}^{2}m_{\phi }^{2}}\left[ v\beta
_{\alpha }-3\alpha +\frac{3}{2}\alpha ^{2}+\frac{m_{\phi }^{2}}{2H^{2}}%
(1-\varsigma ^{2})\right] ,\qquad \mathcal{V}^{\prime }=\frac{2H^{2}}{\hat{%
\kappa}}\left[ v\beta _{\varsigma }-(3-\alpha )\varsigma \right] .
\label{vvp}
\end{equation}

\subsection{Strategy}

We first solve (\ref{vhf}) by assuming generic expansions for the beta
functions $\beta _{\alpha }$ and $\beta _{\varsigma }$. Then we insert the
solutions into (\ref{vvp}) to determine the coefficients of the expansions.

The purpose is to find a solution in parameter space with the right
properties to use perturbative RG methods, that is to say: 1) the beta
functions $\beta _{\alpha }$ and $\beta _{\varsigma }$ are power series in $%
\alpha $ and $\varsigma $; 2) de Sitter limit is the fixed point at $\alpha
=\varsigma =0$; and 3)\ the beta functions behave quadratically around that
point, as in asymptotically free quantum field theories. The existence of a
regime with such properties emerges a posteriori, once the solution is
derived explicitly. We recall that, on the contrary, the regimes mostly
studied in double-field inflation so far rely on numerical methods.

In particular, the expansions of $\beta _{\alpha }$ and $\beta _{\varsigma }$
must start from linear combinations of the monomials $\alpha ^{2}$,$\ \alpha
\varsigma $ and $\varsigma ^{2}$. Moreover, it is natural to demand that the
case $\varsigma =0$ returns the usual single-field Starobinsky inflation,
triggered by the $R^{2}$ term. To this purpose, we can postulate that the
expansion of $\beta _{\varsigma }$ factorizes an overall factor $\varsigma $%
, i.e.%
\begin{equation*}
\beta _{\alpha }=\sum_{n=0}^{\infty }\varsigma ^{n}b_{n}(\alpha ),\qquad
\beta _{\varsigma }=\varsigma \sum_{n=1}^{\infty }\varsigma
^{n-1}c_{n}(\alpha ),
\end{equation*}%
where $b_{n}(\alpha )$ and $c_{n}(\alpha )$ are power series in $\alpha $,
with $b_{0}(\alpha )=\mathcal{O}(\alpha ^{2})$ and $b_{1}(\alpha )=\mathcal{O%
}(\alpha )$, $c_{1}(\alpha )=\mathcal{O}(\alpha )$. Then the second equation
of (\ref{vvp}) implies $\mathcal{V}^{\prime }=0$ at $\varsigma =0$ and the
second equation of (\ref{alfe}) implies $\dot{\varphi}=0$: the scalar $%
\varphi $ sits at the minimum of the potential $\mathcal{V}$ and does not
participate in inflation in this limit.

It is also convenient to expand the solutions for $v$, $h$ and $\varphi $ in
powers of $\varsigma $, the coefficients being functions of $\alpha $:%
\begin{equation}
v(\alpha ,\varsigma )=\sum_{n=0}^{\infty }\varsigma ^{n}v_{n}(\alpha
),\qquad H(\alpha ,\varsigma )=h_{0}(\alpha )\exp \left( \sum_{n=1}^{\infty
}\varsigma ^{n}h_{n}(\alpha )\right) ,\qquad \varphi (\alpha ,\varsigma
)=\sum_{n=1}^{\infty }\varsigma ^{n}\varphi _{n}(\alpha ).  \label{vHph}
\end{equation}

The zeroth-order functions $v_{0}(\alpha )$, $h_{0}(\alpha )$ and $%
b_{0}(\alpha )$ satisfy 
\begin{equation*}
b_{0}v_{0}^{\prime }+v_{0}=1-\alpha ,\qquad v_{0}b_{0}h_{0}^{\prime }=\alpha
h_{0},\qquad b_{0}v_{0}=3\alpha -\frac{3}{2}\alpha ^{2}-\frac{m_{\phi }^{2}}{%
2h_{0}^{2}}.
\end{equation*}%
The first two equations come from (\ref{vhf}) and can be integrated for $%
v_{0}$ and $h_{0}$ by quadratures, given $b_{0}$. Then, the third equation,
which comes from the first equation of (\ref{vvp}), fixes $b_{0}$.\ The
solutions%
\begin{eqnarray}
v_{0}(\alpha ) &=&1-\alpha -2\alpha ^{2}-\frac{29}{3}\alpha ^{3}+\mathcal{O}%
(\alpha ^{4}),  \notag \\
h_{0}(\alpha ) &=&\frac{m_{\phi }}{\sqrt{6\alpha }}\left[ 1-\frac{\alpha }{12%
}+\frac{19}{288}\alpha ^{2}-\frac{373}{3456}\alpha ^{3}+\mathcal{O}(\alpha
^{4})\right] ,  \label{sola} \\
b_{0}(\alpha ) &=&-\alpha ^{2}\left[ 2+\frac{5}{3}\alpha +\frac{56}{9}\alpha
^{2}+\frac{742}{27}\alpha ^{3}+\mathcal{O}(\alpha ^{4})\right] ,  \notag
\end{eqnarray}%
describe the single-field problem ($\varsigma =0$) in the geometric
framework.

The functions $v_{n}(\alpha )$, $h_{n}(\alpha )$ and $\varphi _{n}(\alpha )$
with $n>0$ can be worked out iteratively from equations (\ref{vhf}) by means
of quadratures, given the beta functions. For example, the first functions
satisfy%
\begin{eqnarray*}
&&b_{0}v_{1}^{\prime }+(1+c_{1})v_{1}=-b_{1}v_{0}^{\prime },\qquad \qquad
b_{0}v_{2}^{\prime }+(1+2c_{1})v_{2}=-b_{1}v_{1}^{\prime
}-b_{2}v_{0}^{\prime }-c_{2}v_{1}, \\
&&b_{0}\varphi _{1}^{\prime }+c_{1}\varphi _{1}=-\frac{2}{\hat{\kappa}v_{0}}%
,\qquad \qquad b_{0}\varphi _{2}^{\prime }+2c_{1}\varphi _{2}=-b_{1}\varphi
_{1}^{\prime }-c_{2}\varphi _{1}+\frac{2v_{1}}{\hat{\kappa}v_{0}^{2}}, \\
&&b_{0}h_{1}^{\prime }+c_{1}h_{1}=-\frac{\alpha v_{1}}{v_{0}^{2}}-b_{1}\frac{%
h_{0}^{\prime }}{h_{0}},\qquad b_{0}h_{2}^{\prime
}+2c_{1}h_{2}=-b_{1}h_{1}^{\prime }-b_{2}\frac{h_{0}^{\prime }}{h_{0}}%
-c_{2}h_{1}+\frac{\alpha }{v_{0}^{3}}\left( v_{1}^{2}-v_{2}v_{0}\right) .
\end{eqnarray*}

Assuming that $v_{j}$, $\varphi _{j}$ and $h_{j}$ are known for $j\leqslant
n $, we first find $v_{n+1}$ from equations like those of the first line and 
$\varphi _{n+1}$ from equations like those of the second line. Then we can
work out $h_{n+1}$ from equations like those of the third line.

The solutions can be expanded in powers of $\alpha $, but while $%
v_{n}(\alpha )$ are regular, $\varphi _{n}(\alpha )$ and $h_{n}(\alpha )$
may contain overall negative powers. More explicitly, they have expressions
of the forms%
\begin{equation*}
\varphi _{n}(\alpha ),h_{n}(\alpha )=\frac{1}{\alpha ^{n}}(\text{power
series in }\alpha ),\qquad n>0.
\end{equation*}

Once $v$, $H$ are $\varphi $ are known, $\mathcal{V}$ and $\mathcal{V}%
^{\prime }$ are also known, so (\ref{vvp}) become consistency conditions,
which determine the coefficients of the beta functions $\beta _{\alpha }$
and $\beta _{\varsigma }$. For example, given the potential $\mathcal{V}%
(\varphi )=m^{2}\varphi ^{2}/2$, the first condition coming from the first
formula of (\ref{vvp}) is%
\begin{equation*}
b_{0}v_{1}+b_{1}v_{0}+4b_{0}h_{1}v_{0}-6\alpha (2-\alpha )h_{1}+\frac{%
m_{\phi }^{2}h_{1}}{h_{0}^{2}}=0.
\end{equation*}%
The first two conditions coming from the second formula of (\ref{vvp}) are%
\begin{equation*}
c_{1}v_{0}-3+\alpha -\frac{m^{2}\hat{\kappa}\varphi _{1}}{2h_{0}^{2}}%
=0,\qquad c_{1}(v_{1}+2h_{1}v_{0})+c_{2}v_{0}-2(3-\alpha )h_{1}-\frac{m^{2}%
\hat{\kappa}\varphi _{2}}{2h_{0}^{2}}=0.
\end{equation*}

\subsection{Solution}

The instructions just given are enough to work out the solution, which we
have done up to the orders $\alpha ^{14-p}\varsigma ^{p}$ with $p\leqslant
10 $. We report the result in the variables $\alpha $ and $\lambda $, even
if it is still unclear why we should use them. To the NNLL order, we have%
\begin{eqnarray}
\beta _{\alpha }(\alpha ,\lambda ) &=&-2\alpha ^{2}-\frac{5}{3}\alpha
^{3}+3\alpha ^{2}\lambda \left( \frac{1}{\varrho }-2\right) -\frac{56}{9}%
\alpha ^{4}+\alpha ^{3}\lambda (5-8\varrho )-\frac{9}{\varrho }\alpha
^{2}\lambda ^{2}+\alpha ^{2}\mathcal{O}_{3},  \notag \\
\beta _{\lambda }(\alpha ,\lambda ) &=&2\alpha \lambda (2\varrho -1)+\frac{%
\alpha ^{2}\lambda }{3}\left( 8\varrho ^{2}+2\varrho -3\right) +3\alpha
\lambda ^{2}\left( 4-\frac{1}{\varrho }\right) +\alpha ^{2}\lambda
^{2}\left( 20\varrho -1-\frac{2}{\varrho }\right)  \notag \\
&&+\frac{2\alpha ^{3}\lambda }{9}\left( 16\varrho ^{3}-28\varrho
^{2}+72\varrho -31\right) +\frac{9\alpha \lambda ^{3}}{\varrho }+\alpha
\lambda \mathcal{O}_{3},  \label{bas}
\end{eqnarray}%
where $\mathcal{O}_{n}$ means $\mathcal{O}(\alpha ^{n-m}\lambda ^{m})$, $%
0\leqslant m\leqslant n$. The leading log running couplings are (\ref{llrc}%
). The NLL running couplings, which are needed for the calculations of the
next sections, are derived below.

We see that both positive and negative powers of $\varrho $ appear in the
expansions of the beta functions, so we assume that $\varrho $ is neither
too small nor too large. Moreover, a large $\varrho $ makes the scalar $%
\varphi $ very heavy. In that case, it can be integrated out and effectively
decouples, returning the single-field inflation driven by $\phi $. Also note
that a large $\varrho $ violates the bound (\ref{bond}) of asymptotic de
Sitter freedom.

As far as the functions $v$, $\varphi $ and $H$ are concerned, which are
useful to derive the spectra, we collect their lowest orders in formulas (%
\ref{vfiH}) of the appendix. Observe that%
\begin{eqnarray}
\beta _{\alpha }(\alpha ,\lambda ) &=&\alpha ^{2}\mathcal{A}(\alpha ,\lambda
),\qquad \beta _{\lambda }(\alpha ,\lambda )=\alpha \lambda \mathcal{B}%
(\alpha ,\lambda ),\qquad  \notag \\
\mathcal{V}(\varphi (\alpha ,\lambda )) &=&\frac{\lambda }{\alpha }\mathcal{%
\tilde{V}}(\alpha ,\lambda ),\qquad v(\alpha ,\lambda )=1-\alpha +\alpha
^{2}\Delta v(\alpha ,\lambda ),  \label{properties}
\end{eqnarray}%
where $\mathcal{A}$, $\mathcal{B}$, $\mathcal{\tilde{V}}$ and $\Delta v$ are
power series in $\alpha $ and $\lambda $. In particular, the structures (\ref%
{baw}) are confirmed. We are going to need these properties to show that the
Mukhanov-Sasaki action of the scalar perturbations is perturbative in $%
\alpha $ and $\lambda $.

\subsection{Running couplings}

To the NLL\ order we have%
\begin{eqnarray}
\beta _{\alpha }(\alpha ,\lambda ) &=&-2\alpha ^{2}-\frac{5}{3}\alpha ^{3}+%
\frac{3}{\varrho }\alpha ^{2}\lambda (1-2\varrho ),  \notag \\
\beta _{\lambda }(\alpha ,\lambda ) &=&-2\alpha \lambda (1-2\varrho )+\frac{%
\alpha ^{2}\lambda }{3}\left( 8\varrho ^{2}+2\varrho -3\right) -\frac{3}{%
\varrho }\alpha \lambda ^{2}(1-4\varrho ).  \label{betas}
\end{eqnarray}%
We can work out the running couplings to the same order from the ansatz%
\begin{equation}
\alpha (-\tau )=\frac{\alpha _{k}}{z}\left[ 1+\alpha _{k}f_{1}(z)+\lambda
_{k}f_{2}(z)\right] ,\qquad \lambda (-\tau )=\lambda _{k}z^{2\varrho -1}%
\left[ 1+\alpha _{k}f_{3}(z)+\lambda _{k}f_{4}(z)\right] ,  \label{ansa}
\end{equation}%
where $k$ is a reference momentum scale, $z=1+2\alpha _{k}\ln (-k\tau )$, $%
\alpha _{k}$ and $\lambda _{k}$ denote the couplings at $\tau =-1/k$ and $%
f_{i}(z)$, $i=1,\ldots 4$ denote unknown functions. Inserting (\ref{ansa})
into (\ref{betas}) and working to the required orders, we obtain 
\begin{eqnarray}
\alpha (-\tau ) &=&\frac{\alpha _{k}}{z}\left[ 1-\frac{5\alpha _{k}}{6z}\ln
z-\frac{3\lambda _{k}}{4z\varrho ^{2}}(2\varrho -1)\left( z^{2\varrho
}-1\right) \right] ,  \notag \\
\lambda (-\tau ) &=&\lambda _{k}z^{2\varrho -1}\left\{ 1+(2\varrho -1)\alpha
_{k}\frac{5\ln z+2(z-1)(2\varrho -1)}{6z}\right.  \label{runningc} \\
&&\left. +\frac{3\lambda _{k}}{4\varrho ^{2}z(2\varrho -1)}\left[ (4\varrho
^{2}+2\varrho -1)(z^{2\varrho }-1)+4\varrho (2\varrho ^{2}-4\varrho
+1)\left( z-1\right) \right] \right\} .  \notag
\end{eqnarray}

\subsection{RG invariance of the perturbation spectra}

In the next section we prove that the spectrum $\mathcal{P}_{T}$ of the
tensor fluctuations is RG invariant in the superhorizon limit, where it
satisfies an RG evolution equation of the Callan-Symanzik form, with
vanishing anomalous dimension. As far as the spectra of the scalar
perturbations are concerned, the RG invariant ones can be identified only 
\textit{a posteriori}, using RG invariance as the guiding principle.

Before proceeding, we recall the meaning of RG invariance. Considering the
spectra as functions of $\alpha $, $\lambda $ and $\tau $, they are RG
invariant if they satisfy%
\begin{equation}
\frac{\mathrm{d}\mathcal{P}}{\mathrm{d}\ln |\tau |}=\left( \frac{\partial }{%
\partial \ln |\tau |}+\beta _{\alpha }(\alpha ,\lambda )\frac{\partial }{%
\partial \alpha }+\beta _{\lambda }(\alpha ,\lambda )\frac{\partial }{%
\partial \lambda }\right) \mathcal{P}=0.  \label{RG}
\end{equation}%
If we express $\alpha $ and $\lambda $ as functions of $\tau $, $\alpha _{k}$%
, $\lambda _{k}$, by means of the running couplings, the RG equations imply
that the dependence on $\tau $ drops out and we remain with 
\begin{equation}
\mathcal{P}=\mathcal{\tilde{P}}(\alpha _{k},\lambda _{k}),\qquad \frac{%
\mathrm{d}\mathcal{\tilde{P}}(\alpha _{k},\lambda _{k})}{\mathrm{d}\ln k}%
=-\beta _{\alpha }(\alpha _{k},\lambda _{k})\frac{\partial \mathcal{\tilde{P}%
}(\alpha _{k},\lambda _{k})}{\partial \alpha _{k}}-\beta _{\lambda }(\alpha
_{k},\lambda _{k})\frac{\partial \mathcal{\tilde{P}}(\alpha _{k},\lambda
_{k})}{\partial \lambda _{k}}.  \label{noalfa}
\end{equation}%
This means that the spectra depend on the momentum $k$ only through $\alpha
_{k}$ and $\lambda _{k}$. Finally, expressing $\alpha _{k}$ and $\lambda _{k}
$ as functions of $k/k_{\ast }$ and $\alpha _{\ast }=\alpha (1/k_{\ast })$, $%
\lambda _{\ast }=\lambda (1/k_{\ast })$, where $k_{\ast }$ is the pivot
scale, the RG equation reads%
\begin{equation}
\left( \frac{\partial }{\partial \ln k}+\beta _{\alpha }(\alpha _{\ast
},\lambda _{\ast })\frac{\partial }{\partial \alpha _{\ast }}+\beta
_{\lambda }(\alpha _{\ast },\lambda _{\ast })\frac{\partial }{\partial
\lambda _{\ast }}\right) \mathcal{P}(k/k_{\ast },\alpha _{\ast },\lambda
_{\ast })=0.  \label{RGeq}
\end{equation}

The RG techniques allow us to calculate RG improved power spectra. This
means that, if $\mathcal{P}$ is expanded in powers of $\alpha _{\ast }$ and $%
\lambda _{\ast }$, the products $\alpha _{\ast }\ln (k/k_{\ast })$ and $%
\lambda _{\ast }\ln (k/k_{\ast })$ are considered of order zero and treated
exactly.

\section{Action and fluctuations}

\setcounter{equation}{0}\label{pertu}

In this section we expand the action and give a preliminary discussion about
the physical quantities. Following \cite{ABP}, it is convenient to introduce
an auxiliary field $\Omega $ and write (\ref{Sqgscal}) as%
\begin{eqnarray}
S &=&-\frac{1}{16\pi G}\int \mathrm{d}^{4}x\sqrt{-g}\left[ R+\frac{1}{%
2m_{\chi }^{2}}C_{\mu \nu \rho \sigma }C^{\mu \nu \rho \sigma }-\frac{1}{%
6m_{\phi }^{2}}(2R-\Omega _{0}-\Omega )(\Omega _{0}+\Omega )\right]   \notag
\\
&&+\frac{1}{2}\int \mathrm{d}^{4}x\sqrt{-g}\left( g^{\mu \nu }\partial _{\mu
}\varphi \partial _{\nu }\varphi -m^{2}\varphi ^{2}\right) .  \label{acc}
\end{eqnarray}%
We have also shifted $\Omega $ by an arbitrary function $\Omega _{0}$, to be
determined.

The metric is parametrized as%
\begin{eqnarray}
g_{\mu \nu } &=&\text{diag}(1,-a^{2},-a^{2},-a^{2})+2\text{diag}(\Phi
,a^{2}\Psi ,a^{2}\Psi ,a^{2}\Psi )-\delta _{\mu }^{0}\delta _{\nu
}^{i}\partial _{i}B-\delta _{\mu }^{i}\delta _{\nu }^{0}\partial _{i}B 
\notag \\
&&-2a^{2}\left( u\delta _{\mu }^{1}\delta _{\nu }^{1}-u\delta _{\mu
}^{2}\delta _{\nu }^{2}+v\delta _{\mu }^{1}\delta _{\nu }^{2}+v\delta _{\mu
}^{2}\delta _{\nu }^{1}\right) ,  \label{metr}
\end{eqnarray}%
where $u=u(t,z)$ and $v=v(t,z)$ are the graviton polarizations, chosen, with
no loss of generality, so that their space momentum $\mathbf{k}$ is oriented
along the $z$ axis after Fourier transform.

We expand the matter scalar field $\varphi $ as%
\begin{equation*}
\varphi =\Theta _{0}+\Theta .
\end{equation*}%
where the background function $\Theta _{0}$ is the solution $\varphi (\alpha
,\lambda )$ calculated in the previous section and $\Theta $ is the quantum
fluctuation. See \cite{reviews,defelice,baumann} for reviews on the
parametrizations of the metric fluctuations, their transformations under
diffeomorphisms and common conventions.

The fakeon $\chi _{\mu \nu }$ has spin 2, so it has 5 components, which must
be projected away as anticipated in the introduction. Two of them make the
higher-derivative partners of $u$ and $v$. Another two make the vector
perturbations, which we do not consider since they do not contain physical
polarizations (the fakeon projection trivializes them to the quadratic order 
\cite{ABP}). The fifth, scalar component of $\chi _{\mu \nu }$ is part of
the scalar fluctuations.

The scalar modes are $\Psi $, $\Phi $, $B$, $\Omega $ and $\Theta $. One
such field can be eliminated by means of a gauge choice. Another one appears
algebraically and can be integrated out straightforwardly. A third one can
be identified with the scalar component of $\chi _{\mu \nu }$ just
mentioned. The remaining two are the physical scalar fluctuations, one being
provided by the metric and one by the matter field $\varphi $.

Specifically, we choose the spatially-flat gauge $\Psi =0$ and set%
\begin{equation}
\Omega _{0}=2R_{0}=-6\dot{H}-12H^{2}=6(\alpha -2)H^{2},  \notag
\end{equation}%
where $R_{0}$ is the Ricci scalar calculated on the unperturbed metric. Once
the action (\ref{acc}) is expanded to the quadratic order in the
fluctuations $u$, $v$, $\Phi $, $\Omega $, $\Theta $ and $B$, we see that
the graviton fields $u$ and $v$ decouple from the scalar fields (and from
each other). This allows us to treat the tensor fluctuations and the scalar
fluctuations separately.

The quadratic Lagrangian $\mathcal{L}_{\text{t}}$ of the tensor
perturbations is derived later, see formula (\ref{ltHD}). Here we borrow the
result to analyze it in the superhorizon limit $k/(aH)\rightarrow 0$, where
it becomes 
\begin{equation}
(8\pi G)\frac{\mathcal{L}_{\text{t}}}{a^{3}}=\dot{u}^{2}\left[ 1+(2-\alpha
)H^{2}\left( \frac{2}{m_{\phi }^{2}}+\frac{1}{m_{\chi }^{2}}\right) \right] -%
\frac{\ddot{u}^{2}}{m_{\chi }^{2}}.  \label{ltHD0}
\end{equation}

Clearly, $u=$ constant solves the equations of motion. The arguments of ref. 
\cite{CMBrunning,FakeOnScalar} can then be repeated to show that the $u$
two-point function is RG invariant in the superhorizon limit, and so is the
spectrum of the tensor perturbations. We are going to use this knowledge to
enhance the calculation by one order of magnitude and get to the NNLL order
more easily.

As for the scalar fluctuations, the issue is more involved, due to a mixing
between the physical degrees of freedom. We can define an adiabatic
perturbation $\mathcal{R}_{\text{adiab}}$ and an isocurvature perturbation $%
\mathcal{R}_{\text{iso}}$ (see appendix \ref{adiabiso}). Both $\mathcal{R}_{%
\text{adiab}}$ and $\mathcal{R}_{\text{iso}}$ are gauge invariant, by
construction, and so is any linear combination%
\begin{equation}
\mathcal{R}_{\text{mix}}=C\mathcal{R}_{\text{adiab}}+D\mathcal{R}_{\text{iso}%
},  \label{rmix}
\end{equation}%
where $C$ and $D$ are functions of the background. However, $\mathcal{R}_{%
\text{adiab}}$ and $\mathcal{R}_{\text{iso}}$ are not conserved in the
superhorizon limit. In the approach of this paper, this means that they are
not RG invariant on superhorizon scales. The questions are:\ which
combinations (\ref{rmix})\ are RG\ invariant?\ How to find them?

In single-field inflation, the superhorizon limit of the action shows that $%
\mathcal{R}_{\text{adiab}}$ is automatically RG invariant \cite{CMBrunning},
so we do not need to multiply by a function $C$ of the background. In
double-field inflation it is not simple, in general, to uncover the RG
invariant combinations $\mathcal{R}_{\text{RG}}^{(1)}$ and $\mathcal{R}_{%
\text{RG}}^{(2)}$ from the superhorizon limit of the action.\ Luckily, the
properties of the cosmic RG flow identify $\mathcal{R}_{\text{RG}}^{(1)}$
and $\mathcal{R}_{\text{RG}}^{(2)}$ rather straightforwardly, which
motivates us to propose RG invariance as the guiding principle to identify
the right physical quantities.

In section \ref{scalarocm} we work out the spectra of $\mathcal{R}_{\text{RG}%
}^{(1)}$ and $\mathcal{R}_{\text{RG}}^{(2)}$ to the NLL order by means of
this approach. We upgrade the strategy of \cite{ABP} by using the RG
techniques of \cite{CMBrunning,FakeOnScalar}, which we further extend from
the Einstein frame to the Jordan frame and from single-field inflation to
double-field inflation. We also compare the spectra with those of $\mathcal{R%
}_{\text{adiab}}$ and $\mathcal{R}_{\text{iso}}$.

\section{Tensor perturbations}

\setcounter{equation}{0}\label{tensorQG}

In this section we study the tensor perturbations and work out their
spectrum to the NNLL order. It is easy to show that the quadratic action of
the tensor fluctuations $u$ and $v$ is unaffected by the scalar field $%
\varphi $ and formally coincides with the one of the single-field case found
in \cite{ABP}. After Fourier transforming the space coordinates to the
momentum $\mathbf{k}$, we obtain%
\begin{equation}
(8\pi G)\frac{\mathcal{L}_{\text{t}}}{a^{3}}=\left( \dot{u}^{2}-\frac{%
k^{2}u^{2}}{a^{2}}\right) \left[ 1+2(2-\alpha )\frac{H^{2}}{m_{\phi }^{2}}+%
\frac{k^{2}}{a^{2}m_{\chi }^{2}}\right] +\frac{\dot{u}^{2}}{m_{\chi }^{2}}%
\left[ (2-\alpha )H^{2}+\frac{k^{2}}{a^{2}}\right] -\frac{\ddot{u}^{2}}{%
m_{\chi }^{2}},  \label{ltHD}
\end{equation}%
plus an identical contribution for $v$, where $k=|\mathbf{k}|$. What changes
is that now $H$ and $a$ (and consequently the spectra) depend on both
couplings $\alpha $ and $\lambda $. In (\ref{ltHD}) $\dot{u}^{2}$, $u^{2}$
and $\ddot{u}^{2}$ stand for $\dot{u}_{\mathbf{k}}\dot{u}_{-\mathbf{k}}$, $%
u_{\mathbf{k}}u_{-\mathbf{k}}$ and $\ddot{u}_{\mathbf{k}}\ddot{u}_{-\mathbf{k%
}}$, respectively, where $u_{\mathbf{k}}$ is the Fourier transform of $u$
with respect to the space coordinates. We drop the subscripts $\mathbf{k}$
and $-\mathbf{k}$ when no confusion is expected to arise.

We first eliminate the higher derivatives of $\mathcal{L}_{\text{t}}$ by
adding an auxiliary field $U$ and considering the extended Lagrangian%
\begin{equation}
\mathcal{L}_{\text{t}}^{\prime }=\mathcal{L}_{\text{t}}+\frac{a^{3}}{8\pi
Gm_{\chi }^{2}}\left( m_{\chi }^{2}S-\ddot{u}-f\dot{u}-gu\right) ^{2},
\label{Ltp}
\end{equation}%
where $f$ and $g$ are functions to be determined. The theories described by
the Lagrangians $\mathcal{L}_{\text{t}}$ and $\mathcal{L}_{\text{t}}^{\prime
}$ are equivalent, because $S$ appears algebraically. Indeed, when we solve
the $S$ field equation and insert the solution back into $\mathcal{L}_{\text{%
t}}^{\prime }$, we retrieve $\mathcal{L}_{\text{t}}$. If we keep $S$,
instead, and integrate by parts to remove $\ddot{u}$ from the term
proportional to $\ddot{u}S$, $\mathcal{L}_{\text{t}}^{\prime }$ contains two
fields ($u$ and $S$), but no higher derivatives.

\subsection{De Sitter diagonalization}

We can diagonalize $\mathcal{L}_{\text{t}}^{\prime }$ in the de Sitter limit 
$\alpha =\lambda =0$ by choosing%
\begin{equation*}
f=(3-2\alpha )H,\!\qquad g=\frac{m_{\chi }^{2}}{2}+\frac{k^{2}}{a^{2}}%
-\left( 2\alpha +\frac{\alpha }{\xi }-\alpha ^{2}+v\beta _{\alpha }\right)
H^{2},
\end{equation*}%
and making the field redefinitions%
\begin{eqnarray}
u &=&m_{\phi }\sqrt{\zeta \pi G}\frac{U+V}{aH}\left[ 1+\zeta (\xi -1)\frac{%
\alpha }{2}+\zeta ^{2}(\xi -1)^{2}\frac{\alpha ^{2}}{8}\right] ,  \notag \\
S &=&-\frac{4}{m_{\phi }}\sqrt{\frac{\pi G}{\zeta }}\frac{H}{a}\left[
V+\zeta (\xi -1)\frac{\alpha }{2}U+\zeta ^{2}(\xi -1)^{2}\frac{\alpha ^{2}}{8%
}V\right] ,  \label{uS}
\end{eqnarray}%
where $U$ and $V$ are new fields and $\xi =m_{\phi }^{2}/m_{\chi }^{2}$, $%
\zeta =2/(2+\xi )$. Switching to conformal time, we find the action%
\begin{equation}
S_{\text{t}}=\int \mathrm{d}\tau a\mathcal{L}_{\text{t}}^{\prime }=\int 
\mathrm{d}\tau \left( \mathcal{L}_{U}+\mathcal{L}_{V}+\mathcal{L}%
_{UV}\right) ,  \label{St}
\end{equation}%
where we have separated the $U$ sector $\mathcal{L}_{U}$, from the $V$
sector $\mathcal{L}_{V}$ and the mixing sector $\mathcal{L}_{UV}$.

Since the results to the NNLL order are involved, we just report the
intermediate steps in concise forms, leaving the explicit expressions to the
appendix. We find 
\begin{equation}
\mathcal{L}_{U}=\frac{1}{2}U^{\prime 2}+q_{2}U^{2}+\alpha \mathcal{O}%
_{3},\qquad \mathcal{L}_{V}=-\frac{1}{2}V^{\prime 2}+q_{4}V^{2}+\alpha 
\mathcal{O}_{1},\qquad \mathcal{L}_{UV}=q_{5}UV+q_{6}UV^{\prime }+\alpha 
\mathcal{O}_{3},  \label{diag}
\end{equation}%
where the coefficients $q_{2}$, $q_{4}$, $q_{5}$ and $q_{6}$ are given in
formula (\ref{appa}). The primes on $U$ and $V$ denote derivatives with
respect to $\tau $.

\subsection{Fakeon projection}

The fakeon projection requires to \textquotedblleft integrate out $V$%
\textquotedblright , by first solving the $V$ field equation with the fakeon
prescription and then inserting the solution $V(U)$ back into the action.
The form of $\mathcal{L}_{UV}$ makes it clear that $V(U)$ is $\mathcal{O}%
_{2} $. When $V(U)$ is inserted into (\ref{St}), both $\mathcal{L}_{V}$ and $%
\mathcal{L}_{UV}$ give contributions that are $\mathcal{O}_{4}$. Since our
plan is to work out the spectra to the NNLL order, it is enough to keep the
orders $\mathcal{O}_{3}$, so the projected Lagrangian is just $\mathcal{L}%
_{U}$. Higher-order corrections are more complicated, since $S_{\text{t}}$
contains nonlocalities from $\mathcal{O}_{4}$ onwards.

The projection $V(U)$ makes sense when no tachyonic behaviors are present,
i.e., the ABP bound of ref. \cite{ABP} is fulfilled. Now we show that the
bound is independent of $\alpha $ and $\lambda $ and coincides with the one
of the single-field case, which is $m_{\chi }>m_{\phi }/4$.

The bound is obtained by ensuring that the fakeon projection makes sense not
only in the superhorizon limit $k/(aH)\rightarrow 0$, but also in the
opposite limit $k/(aH)\rightarrow \infty $. As recalled in the introduction,
a crucial feature of primordial cosmology is that in order to make
predictions about the superhorizon limit, it is not enough to study that
limit, but it is necessary to interpolate from the infinite past.

This is also a key ingredient of the fakeon projection, because we must be
sure that $\chi _{\mu \nu }$ is a purely virtual particle at every energy.
Now, fakeons are better understood in flat space, having been introduced for
the theory of scattering in perturbative quantum field theory, but here we
are expanding around a nontrivial background. Luckily, in both limits $%
k/(aH)\rightarrow 0$ and $k/(aH)\rightarrow \infty $ the problem reduces to
a flat-space one, in suitable variables, where the fakeon projection is also
understood \cite{classicization,FLRW}.

Let us consider the $V$ field equation%
\begin{equation}
V^{\prime \prime }+k^{2}V+\frac{4V}{\xi \tau ^{2}}=\mathcal{O}_{2}.
\label{feq0}
\end{equation}%
This form is useful to study the limit $|k\tau |\gg 1$, where $V^{\prime
\prime }+k^{2}V\sim \mathcal{O}_{2}$ and the projection reduces to the
flat-space one in conformal time. The fakeon Green function is given by \cite%
{classicization,FLRW}%
\begin{equation}
\left. \frac{1}{\frac{\mathrm{d}^{2}}{\mathrm{d}\tau ^{2}}+k^{2}}\right\vert
_{\text{f}}=\frac{1}{2k}\sin \left( k|\tau -\tau ^{\prime }|\right) ,
\label{fu}
\end{equation}%
where the subscript \textquotedblleft f\textquotedblright\ means
\textquotedblleft fakeon prescription\textquotedblright .

Introducing the \textquotedblleft inflaton cosmological
time\textquotedblright 
\begin{equation*}
\bar{t}=-\frac{\ln (-H_{0}\tau )}{H_{0}},
\end{equation*}%
where $H_{0}$ is a constant, $\bar{a}=\mathrm{e}^{H_{0}\bar{t}}$ and $W(\bar{%
t})=\bar{a}^{1/2}H_{0}^{2}V$, we can write (\ref{feq0}) in the equivalent
form%
\begin{equation}
\qquad \frac{\mathrm{d}^{2}W}{\mathrm{d}\bar{t}^{2}}+\frac{k^{2}}{\bar{a}^{2}%
}W+\frac{H_{0}^{2}}{4\xi }\left( 16-\xi \right) W=\mathcal{O}_{2},
\label{fequa}
\end{equation}%
which is efficient to study the superhorizon limit $k/(\bar{a}%
H_{0})\rightarrow 0$.

The fakeon Green function $\hat{G}_{\text{f}}(\bar{t},\bar{t}^{\prime })$
for equation (\ref{fequa}) is uniquely determined by requiring that it
matches (\ref{fu}) for $k/(\bar{a}H_{0})\gg 1$. The result is \cite{ABP}%
\begin{equation*}
\hat{G}_{\text{f}}(\bar{t},\bar{t}^{\prime })=\frac{i\pi \mathrm{sgn}(\bar{t}%
-\bar{t}^{\prime })}{4H_{0}\sinh \left( n_{\chi }\pi \right) }\left[
J_{in_{\chi }}(\check{k})J_{-in_{\chi }}(\check{k}^{\prime })-J_{in_{\chi }}(%
\check{k}^{\prime })J_{-in_{\chi }}(\check{k})\right] ,\qquad n_{\chi }=%
\sqrt{\frac{4m_{\chi }^{2}}{m_{\phi }^{2}}-\frac{1}{4}},
\end{equation*}%
where $J_{n}$ denotes the Bessel function of the first kind, $\check{k}=k%
\mathrm{e}^{-H_{0}\bar{t}}/H_{0}$, $\check{k}^{\prime }=k\mathrm{e}^{-H_{0}%
\bar{t}^{\prime }}/H_{0}$, and $\mathrm{sgn}(\bar{t})$ is the sign function.
Thus, the fakeon Green function in the superhorizon limit becomes \cite{ABP}%
\begin{equation}
\left. \frac{1}{\frac{\mathrm{d}^{2}}{\mathrm{d}\bar{t}^{2}}%
+H_{0}^{2}n_{\chi }^{2}}\right\vert _{\text{f}}=\frac{1}{2H_{0}n_{\chi }}%
\sin \left( H_{0}n_{\chi }|\bar{t}-\bar{t}^{\prime }|\right) .  \label{suho}
\end{equation}%
This function exhibits no tachyonic behaviors only for 
\begin{equation}
m_{\chi }>\frac{m_{\phi }}{4}.  \label{ABP}
\end{equation}%
When this bound is violated, we have a hyperbolic sine in (\ref{suho}),
which means that the theory is in a different phase and the large scale
structure of the universe as we know it cannot form. This scenario is beyond
the scope of this paper and needs to be explored separately.

Note that we have not switched from $\tau $ to $t$, which is the
cosmological time of the geometric approach. If we do it, we do not obtain a
harmonic oscillator with constant frequency. Then we need a further
reparametrization to reduce to a problem of that type. An example of such a
reparametrization is precisely the one that switches from $t$ to $\bar{t}$.
Any other one leads to the same conclusion (\ref{ABP}), since a theorem
proved in \cite{ABP} ensures that once the squared mass is constant in some
paramerization, any further reparametrization that leaves it constant
preserves its sign. For this reason, it is enough to derive the ABP\ bound
in the variable $\bar{t}$.

Having established when the projection makes sense, the projected field $%
V(U) $ can be obtained relatively easily, because we just need it in the
superhorizon limit $k/(aH)\rightarrow 0$ \footnote{%
The large $k/(aH)$ behavior is required only for the Bunch-Davies vacuum
condition on the solution of the projected $U$ equation, which, as remarked
above, is unaffected by $V(U)$ to the order we are considering.}. We can
derive $V(U)$ for $k/(aH)\rightarrow 0$ by means of the ansatz%
\begin{equation}
V(U)=s_{1}(\alpha ,\lambda )U+s_{2}(\alpha ,\lambda )\tau U^{\prime },
\label{VU}
\end{equation}%
where $s_{1}$, $s_{2}$ are functions to be determined. We insert this
expression into the $V$ field equation, truncated to the order we are
interested in, which is%
\begin{equation}
V^{\prime \prime }+2q_{4}V+(q_{5}-q_{6}^{\prime })U-q_{6}U^{\prime }=\alpha 
\mathcal{O}_{3}.  \label{Vfie}
\end{equation}%
Then we use the $U$ field equation $U^{\prime \prime }=2q_{2}U+\alpha 
\mathcal{O}_{3}$ to eliminate $U^{\prime \prime }$. Finally, we determine $%
s_{1}$, $s_{2}$ by killing the terms proportional to $U$ and $U^{\prime }$
in (\ref{Vfie}). The result is given in formulas (\ref{s12}) of the appendix.

\subsection{Projected action and RG invariance}

As said, the projected Lagrangian to the order we need is just $\mathcal{L}%
_{U}$. We write the action in the form%
\begin{equation}
S_{\text{t}}^{\text{proj}}=\frac{1}{2}\int \mathrm{d}\tau \left[ U^{\prime
2}-h_{\text{t}}k^{2}U^{2}+\left( 2+\sigma _{\text{t}}\right) \frac{U^{2}}{%
\tau ^{2}}+\mathcal{O}_{4}\right] ,  \label{LUproj}
\end{equation}%
where%
\begin{eqnarray*}
h_{\text{t}} &=&1-2\xi \zeta \alpha ^{2}-\frac{3}{\varrho }\xi \zeta \alpha
^{2}\lambda (2\varrho -1)+\frac{1}{6}\xi \zeta ^{2}\alpha ^{3}\left(
14-11\xi \right) +\mathcal{O}_{4}, \\
\sigma _{\text{t}} &=&9\zeta \alpha ^{2}\left( 1+\frac{3}{2}\lambda \right) +%
\frac{3\zeta ^{2}\alpha ^{3}}{2}(32+37\xi +6\xi ^{2})+\mathcal{O}_{4}.
\end{eqnarray*}%
Defining $\eta =-k\tau $ and $w(\eta )=U(\tau )\sqrt{k}$, (\ref{LUproj})
becomes 
\begin{equation}
S_{\text{t}}^{\text{proj}}=\frac{1}{2}\int \mathrm{d}\eta \left[ w^{\prime
2}-h_{\text{t}}w^{2}+\left( 2+\sigma _{\text{t}}\right) \frac{w^{2}}{\eta
^{2}}\right] ,  \label{mu}
\end{equation}%
where the prime on $w$ stands for the derivative with respect to $\eta $ and
we have dropped the orders that do not contribute to our calculation. The $w$
equation of motion is 
\begin{equation}
w^{\prime \prime }+h_{\text{t}}w-\frac{2+\sigma _{\text{t}}}{\eta ^{2}}w=0.
\label{muck}
\end{equation}

If we want the tensor spectrum to the NNLL order, we actually need the $%
\mathcal{O}_{4}$ contributions to $\sigma _{\text{t}}$. We can infer them
indirectly, by means of RG invariance and the method of \cite{FakeOnScalar}.
What makes this possible is the knowledge, from section \ref{pertu}, that
the tensor mode $u$ is RG invariant in the superhorizon limit\footnote{%
We cannot proceed like this for the scalar perturbations, because the RG
invariant combinations $\mathcal{R}_{\text{RG}}^{(1)}$ and $\mathcal{R}_{%
\text{RG}}^{(2)}$ are not known \textit{a priori}.}.

We parametrize the $\mathcal{O}_{4}$ contributions to $\sigma _{\text{t}}$
as 
\begin{equation}
\zeta ^{2}\alpha ^{2}(c_{1}\alpha ^{2}+c_{2}\alpha \lambda +c_{3}\lambda
^{2})  \label{o4sigmat}
\end{equation}%
and determine the constants $c_{i}$ as follows. First, we decompose $\eta
w(\eta )$ as the sum 
\begin{equation}
\eta w=Q(\ln \eta )+Y(\eta ).  \label{decomp}
\end{equation}%
of a power series $Q(\ln \eta )$ in $\ln \eta $ plus a power series $Y(\eta
) $ in $\eta $ and $\ln \eta $, such that $Y(\eta )\rightarrow 0$
term-by-term for $\eta \rightarrow 0$. Inserting (\ref{decomp}) into (\ref%
{muck}), we find%
\begin{equation}
Q^{\prime \prime }-3Q^{\prime }-\sigma _{\text{t}}Q=\sigma _{\text{t}%
}Y+2\eta Y^{\prime }-\eta ^{2}Y^{\prime \prime }-h_{\text{t}}\eta ^{2}(Y+Q),
\label{QY}
\end{equation}%
where the primes denote the derivatives with respect to the arguments of the
functions. Both sides of equation (\ref{QY}) must be separately zero,
because the left-hand side is an expansion in powers of $\ln \eta $, while
the right-hand side is at least an overall factor $\eta $ times an expansion
in powers of $\ln \eta $. Thus, the $Q$ equation reads%
\begin{equation*}
Q^{\prime \prime }-3Q^{\prime }=\left( \frac{\mathrm{d}}{\mathrm{d}\ln \eta }%
-3\right) \frac{\mathrm{d}Q}{\mathrm{d}\ln \eta }=\sigma _{\text{t}}Q.
\end{equation*}%
Actually, this equation can be further reduced, since the contributions
proportional to $\eta ^{3}$, allowed by the operator in parenthesis, do not
belong to $Q$, by construction. The workaround is to invert that operator
perturbatively, which gives \cite{CMBrunning}%
\begin{equation}
DQ=-\frac{1}{3}\frac{1}{1-\frac{D}{3}}(\sigma _{\text{t}}Q),  \label{DQ}
\end{equation}%
where%
\begin{equation}
D\equiv \frac{\mathrm{d}}{\mathrm{d}\ln \eta }=\beta _{\alpha }\frac{%
\partial }{\partial \alpha }+\beta _{\lambda }\frac{\partial }{\partial
\lambda }.  \label{Dop}
\end{equation}%
It is understood that $\left( 1-\frac{D}{3}\right) ^{-1}$ on the right-hand
side of (\ref{DQ}) must be expanded as a power series in $D$. Indeed, $D$ is 
$\mathcal{O}_{1}$, since the beta functions $\beta _{\alpha }$ and $\beta
_{\lambda }$ are $\mathcal{O}_{2}$ to the leading order.

Viewing $Q(\ln \eta )$ as a function $\tilde{Q}$ of $\alpha $ and $\lambda $%
, we can write 
\begin{equation}
D\tilde{Q}=-\frac{\sigma _{\text{t}}\tilde{Q}}{3}-\frac{1}{9}D(\sigma _{%
\text{t}}\tilde{Q})-\frac{1}{27}D^{2}(\sigma _{\text{t}}\tilde{Q}),
\label{DQt}
\end{equation}%
to the order we need. We express the solution as%
\begin{equation}
\tilde{Q}_{\text{t}}(\alpha ,\lambda ;\alpha _{k},\lambda _{k})=\frac{J_{%
\text{t}}(\alpha ,\lambda )}{J_{\text{t}}(\alpha _{k},\lambda _{k})}\tilde{Q}%
_{\text{t}}(\alpha _{k},\lambda _{k}),  \label{Qtildet}
\end{equation}%
where $J_{\text{t}}$ and $\tilde{Q}_{\text{t}}$ are functions to be
determined. We calculate the constants $c_{i}$ of (\ref{o4sigmat}) from the
RG invariance of $u$ in the superhorizon limit, which ensures that $u$
cannot depend on $\alpha $ and $\lambda $ in that limit.

Specifically, formula (\ref{uS}) gives%
\begin{equation}
u=\frac{m_{\phi }}{aH}\sqrt{\zeta \pi G}\left[ U+s_{1}(\alpha ,\lambda
)U+s_{2}(\alpha ,\lambda )\tau U^{\prime }\right] \left[ 1+\zeta (\xi -1)%
\frac{\alpha }{2}+\zeta ^{2}(\xi -1)^{2}\frac{\alpha ^{2}}{8}\right] .
\label{uU}
\end{equation}%
Replacing $U=w/\sqrt{k}$ with $\tilde{Q}/(\eta \sqrt{k})$, we obtain that
the expression%
\begin{equation}
\left[ J_{\text{t}}+s_{1}(\alpha ,\lambda )J_{\text{t}}+s_{2}(\alpha
,\lambda )(DJ_{\text{t}}-J_{\text{t}})\right] v\left[ 1+\zeta (\xi -1)\frac{%
\alpha }{2}+\zeta ^{2}(\xi -1)^{2}\frac{\alpha ^{2}}{8}\right]  \label{DJt}
\end{equation}%
must be a numerical constant to the NNLL order. Solving (\ref{DQt}) for $J_{%
\text{t}}(\alpha ,\lambda )$ and inserting the solution into (\ref{DJt}), we
find (\ref{Jt}) and (\ref{c3}), the numerical constant being equal to unity.

Note that the expression (\ref{Qtildet}) does not match the one of \cite%
{FakeOnScalar} for $\lambda $, $\lambda _{k}\rightarrow 0$, because the
calculations of \cite{FakeOnScalar} were done in the inflaton framework,
while we are working in the geometric framework here. Only the physical
quantities (spectra, tilts, running coefficients, etc.) need to match, as
they do (see below).

\subsection{Solution of the projected equation of motion}

It remains to determine the constant $\tilde{Q}_{\text{t}}(\alpha
_{k},\lambda _{k})$ of formula (\ref{Qtildet}). To achieve this goal, we
need to solve the equation of motion (\ref{muck}) of the projected action (%
\ref{LUproj}). To impose the correct Bunch-Davies condition, we need to make
a change of variables and eliminate the mass renormalization $h_{\text{t}}$.
Following \cite{CMBrunning}, we rewrite the action (\ref{LUproj}) as%
\begin{equation}
\tilde{S}_{\text{t}}^{\text{prj}}=\frac{1}{2}\int \mathrm{d}\tilde{\eta}%
\left( \tilde{w}^{\prime \hspace{0.01in}2}-\tilde{w}^{2}+\frac{2\tilde{w}^{2}%
}{\tilde{\eta}^{2}}+\tilde{\sigma}_{\text{t}}\frac{\tilde{w}^{2}}{\tilde{\eta%
}^{2}}\right) ,  \label{s2}
\end{equation}%
where the new variable $\tilde{\eta}(\eta )$ is defined as the solution of
the differential equation $\tilde{\eta}^{\prime }(\eta )=\sqrt{h_{\text{t}%
}(\eta )}$ with the initial condition $\tilde{\eta}(0)=0$, while%
\begin{equation}
\tilde{w}(\tilde{\eta}(\eta ))=h_{\text{t}}(\eta )^{1/4}w(\eta ),\qquad 
\tilde{\sigma}_{\text{t}}=\frac{\tilde{\eta}^{2}(\sigma _{\text{t}}+2)}{\eta
^{2}h_{\text{t}}}+\frac{\tilde{\eta}^{2}}{16h_{\text{t}}^{3}}\left( 4h_{%
\text{t}}h_{\text{t}}^{\prime \prime }-5h_{\text{t}}^{\prime \hspace{0.01in}%
2}\right) -2.  \label{wtst}
\end{equation}

The Bunch-Davies vacuum condition for (\ref{s2}) is the usual one,%
\begin{equation}
\tilde{w}(\tilde{\eta})\simeq \frac{\mathrm{e}^{i\tilde{\eta}}}{\sqrt{2}}%
\text{\qquad for }\tilde{\eta}\rightarrow \infty .  \label{bdt}
\end{equation}%
Using (\ref{wtst}) and the running couplings (\ref{runningc}), we expand $%
\tilde{\sigma}_{\text{t}}$ in powers of $\alpha _{k}$ and $\lambda _{k}$,
then expand the $\tilde{w}$ equation of motion by writing%
\begin{equation}
\tilde{w}=\tilde{w}_{0}+\alpha _{k}^{2}\tilde{w}_{2}+\alpha _{k}^{3}\tilde{w}%
_{3}+\alpha _{k}^{2}\lambda _{k}\tilde{w}_{21}+\cdots  \label{expaw}
\end{equation}
The solution is given by 
\begin{eqnarray*}
\tilde{w}_{2}(\tilde{\eta}) &=&\zeta W_{2}(\tilde{\eta}),\qquad \tilde{w}%
_{21}(\tilde{\eta})=\frac{3\zeta }{2}W_{2}(\tilde{\eta}), \\
\tilde{w}_{3}(\tilde{\eta}) &=&\zeta W_{3}(\tilde{\eta})-\frac{\zeta }{6}%
\left[ 4(8+3\xi )-\zeta (32+37\xi +6\xi ^{2})\right] W_{2}(\tilde{\eta}),
\end{eqnarray*}%
the functions $W_{i}(\tilde{\eta})$ being defined in formula (\ref{stada}).
Since each $W_{i}(\tilde{\eta})$, $i>0$, tends to zero for $\tilde{\eta}%
\rightarrow \infty $, (\ref{expaw}) satisfies (\ref{bdt}).

Next, the first formula of (\ref{wtst}) gives $w(\eta )$, where from we can
extract $Q(\ln \eta )$ by means of (\ref{decomp}) and the asymptotic
behaviors (\ref{beha}). At $\ln \eta =0$ we obtain the desired spectral
normalization $\tilde{Q}_{\text{t}}(\alpha _{k},\lambda _{k})=Q(0)$,
reported in formula (\ref{Qwt}). Finally, the $\eta $ dependence of $Q(\ln
\eta )$ provides a check of (\ref{Qtildet}) up to the RG improvement.

\subsection{Perturbation spectrum}

Inserting (\ref{Qwt}) into (\ref{Qtildet}) and using (\ref{decomp}), we
obtain $w(\eta )$ in the superhorizon limit, hence $U(\tau )=w(\eta )/\sqrt{k%
}$ in the same limit. Then (\ref{uU}) gives $u$ and (\ref{ppt}) gives the
spectrum. The result is 
\begin{eqnarray}
\mathcal{P}_{T}\left( k\right) =\frac{4m_{\phi }^{2}\zeta G}{\pi } &&\left[
1-3\zeta \alpha _{k}\left( 1+2\alpha _{k}\gamma _{M}+4\gamma _{M}^{2}\alpha
_{k}^{2}-\frac{\pi ^{2}\alpha _{k}^{2}}{3}-\frac{\alpha _{k}^{2}}{3}\right) +%
\frac{\zeta ^{2}\alpha _{k}^{2}}{8}(94+11\xi )\right.  \notag \\
&&+3\gamma _{M}\zeta ^{2}\alpha _{k}^{3}(14+\xi )-\frac{\zeta ^{3}\alpha
_{k}^{3}}{12}(614+191\xi +23\xi ^{2})-\frac{9}{4\varrho }\zeta \alpha
_{k}\lambda _{k}  \notag \\
&&\left. +\frac{3}{4}\left( \frac{20+\xi }{\varrho }+3(1-2\gamma _{M})(2+\xi
)\right) \zeta ^{2}\alpha _{k}^{2}\lambda _{k}+\mathcal{O}_{4}\right] .
\label{pT}
\end{eqnarray}%
At $\lambda _{k}=0$ it agrees with the one of \cite{FakeOnScalar}, as
expected, but with one caveat: the two couplings $\alpha _{k}$ used here and
there do not coincide, since the single-field calculation of \cite%
{FakeOnScalar} was performed in the inflaton framework (Einstein frame),
while the present calculation is performed in the geometric framework
(Jordan frame). The conversion between the two (check for example the
appendix of \cite{ABP}) is%
\begin{equation}
\alpha _{k}^{\text{infl}}=\alpha _{k}-\frac{\alpha _{k}^{3}}{3}+\mathcal{O}%
(\alpha _{k}^{4}).  \label{scheme}
\end{equation}%
Once the conversion is taken into account, we have a perfect match with the
result of \cite{FakeOnScalar} for $\lambda _{k}=0$.

We see that $\mathcal{P}_{T}\left( k\right) $ has a regular expansion in
powers of $\alpha _{k}$ and $\lambda _{k}$, but also a regular expansion in
powers of $\alpha _{k}$ and $\varsigma _{k}=\sqrt{\alpha _{k}\lambda _{k}}$.
The spectra of the scalar perturbations will only have a regular expansion
in powers of $\alpha _{k}$ and $\lambda _{k}$.

One virtue of the RG approach is that it returns manifestly RG invariant
spectra. As in the dimensional transmutation familiar from quantum field
theory, the sliding scale (which is minus the conformal time $\tau $) has
disappeared from $\mathcal{P}_{T}$, replaced by $1/k$ inside the running
couplings $\alpha _{k}$ and $\lambda _{k}$. The relation (\ref{scheme})\ can
be viewed as a scheme change, the two schemes being the Einstein frame and
the Jordan frame. The match between (\ref{pT}) and the result of \cite%
{FakeOnScalar} at $\lambda _{k}=0$ provides an explicit confirmation that
the physical quantities, such as the spectra, are scheme independent.

The Starobinsky limit is obtained by letting $m_{\chi }$ tend to infinity,
where the theory becomes $R+R^{2}$ plus the extra scalar field $\varphi $.
There we have 
\begin{eqnarray}
\mathcal{P}_{T}^{m_{\chi }=\infty }\left( k\right) &=&\frac{4m_{\phi }^{2}G}{%
\pi }\left[ 1-3\alpha _{k}+\frac{\alpha _{k}^{2}}{4}\left( 47-24\gamma
_{M}\right) \right. -\frac{9\alpha _{k}\lambda _{k}}{4\varrho }  \notag \\
&&-\left( \frac{301}{6}+12\gamma _{M}^{2}-42\gamma _{M}-\pi ^{2}\right)
\alpha _{k}^{3}\left. +\frac{3}{2}\left( \frac{10}{\varrho }+3-6\gamma
_{M}\right) \alpha _{k}^{2}\lambda _{k}+\mathcal{O}_{4}\right] .\quad \qquad
\label{pTStaro}
\end{eqnarray}

\section{Scalar perturbations}

\setcounter{equation}{0}\label{scalarocm}

In this section we study the spectra of the scalar perturbations, which we
compute to the NLL order. We show that, as in the single-field case, they
are unaffected by the fakeon $\chi _{\mu \nu }$, since the first corrections
that depend on $m_{\chi }$ are the NNLL ones. This means that the results of
this section coincide with those of the theory $R+R^{2}$ plus the extra
scalar field $\varphi $.

Nevertheless, to prove the statements just made, we still need to start from
the complete $R+R^{2}+C^{2}$ theory, perform the fakeon projection and
verify that the results are actually $m_{\chi }$ independent to the NLL
order. We do so by expanding in powers of $\xi =m_{\phi }^{2}/m_{\chi }^{2}$.

The four scalar fields $\Omega ,B,\Phi ,\Theta $ appearing in the action
play the following roles: $\Phi $ is an auxiliary field and can be
integrated out straightforwardly, $B$ (plus certain corrections) is the
scalar fakeon, while $\Omega $ and $\Theta $ (plus corrections) encode the
physical perturbations $U_{1}$ and $U_{2}$. The nontrivial mixing between $%
U_{1}$ and $U_{2}$ complicates the identification of the RG invariant
combinations. The difficulty can be overcome if we use RG invariance as a
guiding principle. Since most formulas we are going to meet in the
intermediate steps are lengthy, we cannot report them here, but just give
the main details to work out the final result.

We expand the Lagrangian (\ref{acc}) by means of (\ref{metr}) (with $u=v=0$%
)\ to the quadratic order in the fluctuations. Then we make the field
redefinitions $\Omega ,B,\Phi ,\Theta \rightarrow U_{1},E,\tilde{\Phi},U_{2}$%
, where%
\begin{equation}
\Omega =8m_{\phi }\sqrt{3\pi G}\frac{H}{a}U_{1},\qquad B=2\sqrt{3\pi G}\frac{%
a}{k^{2}}E,\qquad \Phi =8m_{\phi }\sqrt{3\pi G}\frac{\tilde{\Phi}}{aH}%
,\qquad \Theta =\frac{U_{2}}{a}.  \label{psib}
\end{equation}%
Next, we simplify the Lagrangian by using the equations (\ref{fried}) and (%
\ref{vvp}) of the background metric and the background scalar $\varphi $. At
the end, we switch to conformal time. Using the properties (\ref{properties}%
), it is easy to show that the quadratic Lagrangian we obtain is analytic in 
$\alpha $ and $\lambda $, apart from the linear terms in $U_{2}$ and $%
U_{2}^{\prime }$, which are multiplied by an extra factor $\sqrt{\lambda }$.
In particular, the de Sitter limit $\alpha ,\lambda \rightarrow 0$ is
regular and reads%
\begin{eqnarray*}
\mathcal{L}_{\text{dS}} &=&\frac{1}{2}U_{2}^{\prime 2}-\frac{k^{2}}{2}%
U_{2}^{2}+\frac{U_{2}^{2}}{\tau ^{2}}-\frac{E^{\prime 2}}{2m_{\chi }^{2}}+%
\frac{2U_{1}E^{\prime }}{\tau m_{\phi }}-\frac{6EU_{1}}{\tau ^{2}m_{\phi }}-%
\frac{2}{\tau ^{2}}U_{1}^{2} \\
&&+\frac{4\tilde{\Phi}}{\tau ^{2}}\left( 12U_{1}+12\frac{E}{m_{\phi }}+6\tau
U_{1}^{\prime }+k^{2}\tau ^{3}\frac{m_{\phi }}{m_{\chi }^{2}}E^{\prime
}-2k^{2}\tau ^{2}U_{1}\right) -\frac{8\tilde{\Phi}^{2}}{\tau ^{2}}\left( 36+%
\frac{m_{\phi }^{2}}{m_{\chi }^{2}}k^{4}\tau ^{4}\right) .
\end{eqnarray*}%
In this limit the extra scalar $U_{2}$ decouples and the quantum gravity
sector can be treated as in the single-scalar case. The negative sign in
front of the $E$ kinetic term $E^{\prime 2}$ shows that $E$ (plus
corrections) is the fakeon.

\subsection{ABP\ bound}

If we integrate out the auxiliary field $\tilde{\Phi}$, we obtain, on
superhorizon scales $k/(aH)\rightarrow 0$ in the de Sitter limit,%
\begin{equation*}
\mathcal{L}_{\text{dS}}^{k\rightarrow 0}=\frac{1}{2}U_{1}^{\prime 2}+\frac{%
U_{1}^{2}}{\tau ^{2}}+\frac{1}{2}U_{2}^{\prime 2}+\frac{U_{2}^{2}}{\tau ^{2}}%
-\frac{1}{2m_{\chi }^{2}}\left( E^{\prime 2}-4\frac{m_{\chi }^{2}}{m_{\phi
}^{2}}\frac{E^{2}}{\tau ^{2}}\right) .
\end{equation*}%
The $E$-dependent part of the action reads, in cosmological time,%
\begin{equation*}
\int \mathrm{d}\tau \mathcal{L}_{C\text{dS}}^{k\rightarrow 0}\simeq -\frac{1%
}{2m_{\chi }^{2}}\int a\mathrm{d}t\left[ \dot{E}^{2}-\frac{2m_{\chi }^{2}}{%
3\alpha }E^{2}\right] ,
\end{equation*}%
up to subleading corrections, where we have used $v=-1/(aH\tau )=1+\mathcal{O%
}_{1}$ and $H=m_{\phi }/\sqrt{6\alpha }(1+\mathcal{O}_{1})$. As in the
previous section, we can read the ABP bound \cite{ABP} by defining a new
cosmological time $t^{\prime }=t^{\prime }(t)$ to reduce the problem to one
with constant coefficients, up to corrections of higher orders. For example,
we can choose $\mathrm{d}t/\mathrm{d}t^{\prime }=\sqrt{\alpha }$ and $\tilde{%
E}(t^{\prime })=\alpha ^{-1/4}E(t)\sqrt{a}$, which gives%
\begin{equation*}
\int \mathrm{d}\tau \mathcal{L}_{C\text{dS}}^{k\rightarrow 0}\simeq -\frac{1%
}{2m_{\chi }^{2}}\int \mathrm{d}t^{\prime }\left[ \left( \frac{\mathrm{d}%
\tilde{E}}{\mathrm{d}t^{\prime }}\right) ^{2}+\left( m_{\phi }^{2}-16m_{\chi
}^{2}\right) \frac{\tilde{E}^{2}}{24}\right] .
\end{equation*}%
The squared mass of the fakeon is positive if and only if (\ref{ABP}) holds,
so the ABP\ bound of the scalar perturbations coincides with the one of the
tensor perturbations. Again, any reparametrization that gives a constant
squared mass leads to the same ABP\ bound. These properties suggest that the
ABP\ bound is a universal property of the theory.

At this point, we can proceed to project the fakeon out.

\subsection{Fakeon projection}

When we restore the dependence on $\alpha $ and $\lambda $, we move away
from the de Sitter point. To study this situation, we still eliminate the
auxiliary field $\tilde{\Phi}$ and then expand in powers of $\xi $ to
perform the fakeon projection, which expresses $E$ in terms of $U_{1}$, $%
U_{2}$ and their derivatives. We have done this explicitly to the order $\xi
^{3}$ included. In the small $\xi $ expansion, the projection is
straightforward, since the $E$ field equation has the form%
\begin{equation}
\frac{4\hat{E}}{\xi \tau ^{2}}-\frac{k^{2}}{3}\hat{E}+\hat{E}^{\prime \prime
}=\mathcal{O}_{1}+\mathcal{O}(\xi ),  \label{Csol}
\end{equation}%
having defined 
\begin{equation*}
E=\sqrt{\frac{3}{\pi G}}\hat{E}+\frac{m_{\phi }k^{2}\tau ^{2}}{6}U_{1},
\end{equation*}%
so the solution can be worked out algebraically by iteration.

Inserting the solution back into the Lagrangian, we obtain the projected
Lagrangian, which has the form%
\begin{equation}
\mathcal{L}_{\text{MS}}=\frac{1}{2}\sum_{i=1}^{2}\left[ U_{i}^{\prime
2}+\left( \frac{2}{\tau ^{2}}-k^{2}\right) U_{i}^{2}\right] +\mathcal{O}_{1}+%
\mathcal{O}(\xi ).  \label{hdscal}
\end{equation}%
The corrections can contain higher derivatives of $U_{1}$, multiplied by
higher powers of $\xi $. They can be eliminated by means of a change of
variables%
\begin{equation}
U_{1}=u_{1}\left( 1-\frac{\alpha }{2}+\frac{\alpha ^{2}}{2}+\frac{9}{%
8\varrho }\alpha \lambda \right) -\alpha \sqrt{\frac{3\lambda }{2}}u_{2}+%
\mathcal{O}_{5/2}+\mathcal{O}(\xi ),\qquad U_{2}=u_{2},  \label{Uu}
\end{equation}%
which turns (\ref{hdscal}) into the standard form%
\begin{eqnarray}
\mathcal{L}_{\text{MS}} &=&\frac{1}{2}\sum_{i=1}^{2}\left[ u_{i}^{\prime
2}+\left( \frac{2}{\tau ^{2}}-k^{2}\right) u_{i}^{2}\right] +\frac{\alpha }{%
\tau }\sqrt{6\lambda }u_{1}^{\prime }u_{2}+\frac{\alpha }{\tau ^{2}}\left[
(3+10\alpha )u_{1}^{2}-\frac{9}{2\varrho }\lambda u_{1}^{2}\right]  \notag \\
&&+\frac{\alpha u_{2}}{\tau ^{2}}\left[ \frac{3}{2}(1-2\varrho )(1+4\alpha
)u_{2}+\frac{11}{2}\varrho \alpha u_{2}+4\sqrt{6\lambda }u_{1}-\frac{9}{2}%
\lambda u_{2}\right] +\mathcal{O}_{5/2}.  \label{lau}
\end{eqnarray}%
Note that this expression is exact in $\xi $, which proves our statement:
the scalar spectra are unaffected by the fakeon to the NLL order.

\subsection{Mukhanov-Sasaki equations}

Having two mixing scalars, we need to work with vectors and matrices. Let us
define%
\begin{eqnarray}
&&w(\eta )=\sqrt{k}\left( 
\begin{tabular}{l}
$u_{1}$ \\ 
$u_{2}$%
\end{tabular}%
\right) ,\qquad \Xi =\alpha \sqrt{\frac{3\lambda }{2}}\left( 
\begin{tabular}{cc}
$0$ & $-1$ \\ 
$1$ & $0$%
\end{tabular}%
\right) +\left( \!%
\begin{tabular}{cc}
$\mathcal{O}_{3}$ & $\mathcal{O}_{5/2}$ \\ 
$\mathcal{O}_{5/2}$ & $\mathcal{O}_{3}$%
\end{tabular}%
\!\right) ,  \notag \\
\Sigma &=&\alpha \left( 
\begin{tabular}{cc}
$6+22\alpha -\frac{9\lambda }{\varrho }$ & $\sqrt{\frac{3\lambda }{2}}\left(
9+(3-2\varrho )\alpha \right) $ \\ 
$\sqrt{\frac{3\lambda }{2}}\left( 9+(3-2\varrho )\alpha \right) $ & $%
3(1-2\varrho )+(12-13\varrho )\alpha -9\lambda $%
\end{tabular}%
\right) +\left( \!%
\begin{tabular}{cc}
$\mathcal{O}_{3}$ & $\mathcal{O}_{5/2}$ \\ 
$\mathcal{O}_{5/2}$ & $\mathcal{O}_{3}$%
\end{tabular}%
\!\right) \!.\qquad  \label{wu}
\end{eqnarray}%
The Mukhanov-Sasaki action is%
\begin{equation}
S_{\text{MS}}=\int \mathrm{d}\eta \hspace{0.01in}\mathcal{L}_{\text{MS}}=%
\frac{1}{2}\int \mathrm{d}\eta \left[ w^{\prime \text{T}}w^{\prime }-w^{%
\text{T}}w+\frac{2}{\eta ^{2}}w^{\text{T}}w+\frac{1}{\eta ^{2}}w^{\text{T}%
}\Sigma w+\frac{2}{\eta }w^{\text{T}}\Xi w^{\prime }\right] ,
\label{muchmatrac}
\end{equation}%
where the superscript T means transpose. The MS equations are%
\begin{equation}
w^{\prime \prime }+w-\frac{2}{\eta ^{2}}w=\frac{1}{\eta ^{2}}\left( \Sigma
-\Xi +D\Xi \right) w+\frac{2}{\eta }\Xi w^{\prime }.  \label{muchmatr}
\end{equation}

\subsection{RG\ invariance}

To study RG invariance we write 
\begin{equation}
\eta w(\eta )=Q(\ln \eta )+Y(\eta ),  \label{decompomatr}
\end{equation}%
where $Q(\ln \eta )$ is a vector with entries that are power series in $\ln
\eta $, while $Y(\eta )$ is a vector with entries that are power series in $%
\eta $ and $\ln \eta $ and such that $Y(\eta )\rightarrow 0$ term-by-term
for $\eta \rightarrow 0$. Proceeding as in \cite{CMBrunning} and above
formula (\ref{DQ}) (but paying attention to the fact that now we have to
deal with matrices and vectors), it is easy to derive the $Q$ equations%
\begin{equation*}
DQ=-\frac{1}{3}\frac{1}{1-\frac{D}{3}}\left[ \left( \Sigma -3\Xi -D\Xi
\right) Q+2D(\Xi Q)\right] .
\end{equation*}%
As usual, the right-hand side must be expanded in powers of $D$. Since the
beta functions are $\mathcal{O}_{2}$ to the lowest order, $D$ is $\mathcal{O}%
_{1}$, while $\Sigma $ is also $\mathcal{O}_{1}$ and $\Xi $ is $\mathcal{O}%
_{3/2}$. To the NLL order we can truncate the equation to%
\begin{equation}
D\tilde{Q}=-\frac{1}{3}\tilde{\Sigma}\tilde{Q},  \label{DQtilde}
\end{equation}%
where%
\begin{equation*}
\tilde{Q}=\left( 1+\frac{\Sigma }{9}+\frac{\Xi }{3}\right) Q,\qquad \tilde{%
\Sigma}=(\Sigma -3\Xi -D\Xi )\left( 1+\frac{\Sigma }{9}+\frac{\Xi }{3}%
\right) ^{-1}.
\end{equation*}%
The solution is%
\begin{equation}
\tilde{Q}(\ln \eta )=K(\eta )\tilde{Q}_{0},  \label{Qsol}
\end{equation}%
where $\tilde{Q}_{0}$ is a constant vector and the kernel $K$ is given by
the ordered exponential%
\begin{eqnarray}
K(\eta ) &=&T\exp \left( -\frac{1}{3}\int_{0}^{\ln \eta }\mathrm{d}\ln \eta
^{\prime }\hspace{0.01in}\tilde{\Sigma}(\eta ^{\prime })\right)  \notag \\
&\equiv &1-\frac{1}{3}\int_{0}^{\ln \eta }\mathrm{d}\ln \eta ^{\prime }%
\hspace{0.01in}\hspace{0.01in}\tilde{\Sigma}(\eta ^{\prime })+\frac{1}{9}%
\int_{0}^{\ln \eta }\mathrm{d}\ln \eta ^{\prime }\hspace{0.01in}\hspace{%
0.01in}\tilde{\Sigma}(\eta ^{\prime })\int_{0}^{\ln \eta ^{\prime }}\mathrm{d%
}\ln \eta ^{\prime \prime }\hspace{0.01in}\hspace{0.01in}\tilde{\Sigma}(\eta
^{\prime \prime })+\cdots .  \label{Oexp}
\end{eqnarray}

It is always possible to write $\tilde{Q}(\ln \eta )$ as the product $%
\mathcal{K}(\alpha ,\lambda )X_{0}$ of a matrix $\mathcal{K}$ that depends
only on $\alpha $ and $\lambda $, times an arbitrary constant vector $X_{0}$%
. To prove this statement, we use the running couplings (\ref{runningc}) to,
say, express $\lambda $ as a function 
\begin{equation}
\lambda =f_{\lambda }(\alpha ,\lambda _{0})  \label{ansat}
\end{equation}%
of $\alpha $ and an arbitrary constant $\lambda _{0}$. The function $%
f_{\lambda }$ is the solution of the first order differential equation%
\begin{equation*}
\frac{\mathrm{d}f_{\lambda }}{\mathrm{d}\alpha }=\frac{\beta _{\lambda
}(\alpha ,f_{\lambda })}{\beta _{\alpha }(\alpha ,f_{\lambda })}.
\end{equation*}%
obtained by inserting (\ref{ansat}) into $\beta _{\lambda }(\alpha ,\lambda
)=\mathrm{d}\lambda /\mathrm{d}\ln \eta $, and $\lambda _{0}$ is the initial
condition. Then, we can write the differential operator $D$ appearing in (%
\ref{DQtilde}) as%
\begin{equation*}
D=\frac{\mathrm{d}}{\mathrm{d}\ln \eta }=\beta _{\alpha }\frac{\partial }{%
\partial \alpha }+\beta _{\lambda }\frac{\partial }{\partial \lambda }=\beta
_{\alpha }(\alpha ,f_{\lambda })\frac{\mathrm{d}}{\mathrm{d}\alpha },
\end{equation*}%
where the total derivative $\mathrm{d}/\mathrm{d}\alpha $ also acts on the $%
\alpha $ dependence due to $f_{\lambda }$.

Since the initial conditions for the system of differential equations (\ref%
{DQtilde}) are collected into a constant vector $X_{0}$ on the right of
everything, the solution can be written as the product $\mathcal{\tilde{K}}%
(\alpha ,\lambda _{0})X_{0}$ of a matrix $\mathcal{\tilde{K}}(\alpha
,\lambda _{0})$ depending on $\alpha $ and $\lambda _{0}$ times $X_{0}$.
Inverting (\ref{ansat}), $\lambda _{0}$ can be expressed as a function of $%
\alpha $ and $\lambda $. This allows us to write $\mathcal{\tilde{K}}(\alpha
,\lambda _{0})$ as a matrix $\mathcal{K}(\alpha ,\lambda )$ whose entries
are functions of $\alpha $ and $\lambda $, as we wished to prove.

At the practical level, we evaluate the solution $\tilde{Q}$ as follows.
First, we use the running couplings of formulas (\ref{runningc}) to expand $%
\Sigma $ and $\Xi $ in powers of $\alpha _{k}$ and $\lambda _{k}$. This
makes the dependence of $\ln \eta $ explicit and allows us to perform the
integrals of (\ref{Oexp}). From (\ref{Qsol}) we learn that $\tilde{Q}(\ln
\eta )=K(\eta )\tilde{Q}_{0}=\mathcal{K}(\alpha ,\lambda )X_{0}$, which
implies $\tilde{Q}_{0}=\mathcal{K}(\alpha _{k},\lambda _{k})X_{0}$ at $\ln
\eta =0$. Since $X_{0}$ is arbitrary, we can rewrite the result of (\ref%
{Oexp}) in the factorized form%
\begin{equation*}
K(\eta )=\mathcal{K}(\alpha ,\lambda )\mathcal{K}^{-1}(\alpha _{k},\lambda
_{k}).
\end{equation*}%
We find 
\begin{equation*}
\mathcal{K}(\alpha ,\lambda )=\left( 
\begin{tabular}{cc}
$\alpha \left( 1+\frac{13}{6}\alpha \right) +\mathcal{O}_{3}$ & $\qquad -%
\frac{1}{\varrho }\sqrt{\frac{3\lambda }{2}}\left( \lambda +\mathcal{O}%
_{2}\right) $ \\ 
$\sqrt{\frac{3\lambda }{2}}\left( \alpha +\mathcal{O}_{2}\right) $ & $\qquad
\lambda \left[ 1+\frac{19-16\varrho }{12}\alpha -\frac{3(3-2\varrho )}{%
4\varrho (1-2\varrho )}\lambda \right] +\mathcal{O}_{3}$%
\end{tabular}%
\right) \left( 
\begin{tabular}{cc}
$1$ & $0$ \\ 
$0$ & $\frac{1}{\sqrt{\lambda }}$%
\end{tabular}%
\right) .
\end{equation*}%
The further factorization appearing here is introduced to better organize
the orders we are neglecting.

We conclude that the solutions of the Mukhanov-Sasaki equations (\ref%
{muchmatr}) read, in the superhorizon limit,%
\begin{equation}
w=\sqrt{k}\left( 
\begin{tabular}{l}
$u_{1}$ \\ 
$u_{2}$%
\end{tabular}%
\right) \simeq \frac{1}{\eta }\left( 1+\frac{\Sigma }{9}+\frac{\Xi }{3}%
\right) ^{-1}\mathcal{K}(\alpha ,\lambda )\mathcal{K}^{-1}(\alpha
_{k},\lambda _{k})\tilde{Q}_{0}.  \label{suso}
\end{equation}%
At this point, what remains to do is compute $\tilde{Q}_{0}$.

\subsection{Solution of the Mukhanov-Sasaki equations}

To find $\tilde{Q}_{0}$, we go back to equations (\ref{muchmatr}) and study
them beyond the superhorizon limit. We need to impose the Bunch-Davies
vacuum condition for large momenta $k/(aH)\gg 1$. Equations (\ref{muchmatr})
contain a term proportional to $w^{\prime }/\eta $, which turns out to be
incompatible with such a condition. We remedy by switching to new variables $%
\varpi $ through the field redefinition%
\begin{equation}
w=\varpi +\Xi \varpi \ln \eta .  \label{Wt}
\end{equation}%
The action becomes%
\begin{equation}
S_{\text{MS}}=\int \mathrm{d}\eta \hspace{0.01in}\mathcal{L}_{\text{MS}}=%
\frac{1}{2}\int \mathrm{d}\eta \left[ \varpi _{-\mathbf{k}}^{\prime \text{T}%
}\varpi _{\mathbf{k}}^{\prime }-\varpi _{-\mathbf{k}}^{\text{T}}\varpi _{%
\mathbf{k}}+\frac{1}{\eta ^{2}}\varpi _{-\mathbf{k}}^{\text{T}}(2+\Sigma
)\varpi _{\mathbf{k}}\right] ,  \label{SMSt}
\end{equation}%
plus higher orders, having restored the subscripts $\mathbf{k}$ and $-%
\mathbf{k}$. The momenta are%
\begin{equation}
\Pi _{-\mathbf{k}}=-\frac{\delta \mathcal{L}_{\text{MS}}}{\delta \varpi _{-%
\mathbf{k}}^{\prime \text{T}}}=-\varpi _{\mathbf{k}}^{\prime },  \label{Pi}
\end{equation}%
the minus sign being due to the fact that the orientation of $\eta $ (which
is equal to $-k\tau $) is opposite to the orientation of time. We quantize
the system by setting 
\begin{equation}
\lbrack \hat{\Pi}_{\mathbf{k}},\hat{\varpi}_{\mathbf{k}^{\prime }}^{\text{T}%
}]=-i\mathbb{I}(2\pi )^{3}\delta ^{(3)}(\mathbf{k}-\mathbf{k}^{\prime
}),\qquad \lbrack \hat{\varpi}_{\mathbf{k}},\hat{\varpi}_{\mathbf{k}^{\prime
}}^{\text{T}}]=[\hat{\Pi}_{\mathbf{k}},\hat{\Pi}_{\mathbf{k}^{\prime }}^{%
\text{T}}]=0,  \label{q1}
\end{equation}%
where $\hat{\Pi}_{\mathbf{k}}$ and $\hat{\varpi}_{\mathbf{k}}$ are the
operators associated with $\Pi _{\mathbf{k}}$ and $\varpi _{\mathbf{k}}$ and 
$\mathbb{I}$ is the 2$\times $2 identity matrix. Then we write%
\begin{equation*}
\hat{\varpi}_{\mathbf{k}}=P(\eta )\hat{A}_{\mathbf{k}}+P^{\ast }(\eta )\hat{A%
}_{-\mathbf{k}}^{\dag },\qquad \hat{\Pi}_{\mathbf{k}}=-P^{\prime }(\eta )%
\hat{A}_{-\mathbf{k}}-P^{\prime \ast }(\eta )\hat{A}_{\mathbf{k}}^{\dag },
\end{equation*}%
where $P(\eta )$ is an $\eta $ dependent two-by-two matrix, to be
determined, while 
\begin{equation*}
\hat{A}_{\mathbf{k}}=\left( 
\begin{tabular}{l}
$\hat{a}_{1\mathbf{k}}$ \\ 
$\hat{a}_{2\mathbf{k}}$%
\end{tabular}%
\right) ,\qquad \hat{A}_{\mathbf{k}}^{\dag }=\left( 
\begin{tabular}{l}
$\hat{a}_{1\mathbf{k}}^{\dag }$ \\ 
$\hat{a}_{2\mathbf{k}}^{\dag }$%
\end{tabular}%
\right) ,
\end{equation*}%
are vectors of ($\eta $-independent) annihilation and creation operators $%
\hat{a}_{i\mathbf{k}}$, $\hat{a}_{i\mathbf{k}}^{\dag }$, satisfying%
\begin{equation}
\lbrack \hat{A}_{\mathbf{k}},\hat{A}_{\mathbf{k}^{\prime }}^{\dag \text{T}}]=%
\mathbb{I}(2\pi )^{3}\delta ^{(3)}(\mathbf{k}-\mathbf{k}^{\prime }),\qquad
\lbrack \hat{A}_{\mathbf{k}},\hat{A}_{\mathbf{k}^{\prime }}^{\text{T}}]=[%
\hat{A}_{\mathbf{k}}^{\dag },\hat{A}_{\mathbf{k}^{\prime }}^{\dag \text{T}%
}]=0.  \label{q2}
\end{equation}%
Then, (\ref{q1}) and (\ref{q2}) give%
\begin{equation*}
P^{\prime }P^{\ast \text{T}}-P^{\ast \prime }P^{\text{T}}=i\mathbb{I}.
\end{equation*}

We proceed by expanding%
\begin{equation}
P=P_{0}+\alpha _{k}P_{1}+\alpha _{k}\sqrt{\lambda _{k}}P_{2}+\alpha
_{k}^{2}P_{3}+\alpha _{k}\lambda _{k}P_{4}+\mathcal{O}_{5/2}.  \label{Utilde}
\end{equation}%
where the $P_{i}$ are matrices of functions of $\eta $. We expand (the
operatorial versions of) the MS equations derived from (\ref{SMSt}) and
derive the equations solved by each $P_{i}$. The solutions are reported in
formula (\ref{scalauto}) of appendix \ref{appscal} to the orders we need. It
is easy to check that the Bunch-Davies conditions%
\begin{equation*}
P\simeq \frac{\mathrm{e}^{i\eta }}{\sqrt{2}}\mathbb{I}\text{\qquad for }\eta
\gg 1
\end{equation*}%
are satisfied.

In the superhorizon limit we use the asymptotic behaviors (\ref{beha}) and
find the operatorial version of formula (\ref{suso}) with $\hat{w}_{\mathbf{k%
}}$ on the left-hand side and 
\begin{equation}
\tilde{Q}_{0}=\frac{i}{\sqrt{2}}\left[ \mathbb{I}+\frac{\alpha _{k}}{3}%
\left( 7-3\tilde{\gamma}_{M}\right) \left( \!%
\begin{tabular}{cc}
$2$ & $0$ \\ 
$0$ & $1-2\varrho $%
\end{tabular}%
\!\right) +\left( \!%
\begin{tabular}{cc}
$\mathcal{O}_{2}$ & $\mathcal{O}_{3/2}$ \\ 
$\mathcal{O}_{3/2}$ & $\mathcal{O}_{2}$%
\end{tabular}%
\!\right) \right] (\hat{A}_{\mathbf{k}}-\hat{A}_{-\mathbf{k}}^{\dag })
\label{Qt0}
\end{equation}%
on the right-hand side.

\subsection{Perturbation spectra}

The matrix $\mathcal{R}_{\text{RG}}$ of RG invariant perturbations is
obtained by multiplying both sides of equation (\ref{suso}) by\ $\eta $
times an $\alpha ,\lambda $-dependent matrix, to remove any dependence on $%
\tau $ in the superhorizon limit. This leaves us with an expression that
just depends on $\alpha _{k}$ and $\lambda _{k}$, which reads%
\begin{equation}
\mathcal{R}_{\text{RG}}=\frac{C}{k^{3/2}}\mathcal{K}^{-1}(\alpha ,\lambda
)\left( 1+\frac{\Sigma }{9}+\frac{\Xi }{3}\right) \eta \hat{w}_{\mathbf{k}%
}\simeq \frac{C}{k^{3/2}}\mathcal{K}^{-1}(\alpha _{k},\lambda _{k})\tilde{Q}%
_{0},  \label{RGcomb}
\end{equation}%
where $C$ is a constant matrix. The factor $k^{-3/2}$ is determined by the
behavior of the curvature perturbations $\mathcal{R}_{\text{adiab}}$ and $%
\mathcal{R}_{\text{iso}}$ of formulas (\ref{adiso}) in the de Sitter limit.

The matrix $C$ cannot be fixed by RG invariance, but we can arrange it so
that one entry of the matrix $\mathcal{R}_{\text{RG}}$ matches the curvature
perturbation $\mathcal{R}_{\text{adiab}}$ in the de Sitter limit. The
behaviors of the other entries of $\mathcal{R}_{\text{RG}}$ do not allow us
to match $\mathcal{R}_{\text{iso}}$.

Specifically, we choose $C$ proportional to the identity matrix and
determine the overall constant so that the 1-1 entry of the matrix $\mathcal{%
R}_{\text{RG}}$ matches $\mathcal{R}_{\text{adiab}}$ for $\alpha
_{k},\lambda _{k}\simeq 0$. Using (\ref{Qt0}), we find 
\begin{equation*}
\mathcal{R}_{\text{RG}}=\sqrt{\frac{\pi G}{6}}\frac{im_{\phi }}{k^{3/2}}T%
\left[ \left( 
\begin{tabular}{cc}
$\frac{1}{\alpha _{k}}\mathcal{R}_{\text{RG}}^{(11)}+\mathcal{O}_{1}$ & $%
\qquad \frac{1}{\alpha _{k}\varrho }\sqrt{\frac{3\lambda _{k}}{2}}+\mathcal{O%
}_{1/2}$ \\ 
$-\sqrt{\frac{3}{2\lambda _{k}}}+\mathcal{O}_{1/2}$ & \qquad $\frac{1}{%
\lambda _{k}}\mathcal{R}_{\text{RG}}^{(22)}+\mathcal{O}_{1}$%
\end{tabular}%
\right) \right] (\hat{A}_{\mathbf{k}}-\hat{A}_{-\mathbf{k}}^{\dag }),
\end{equation*}%
where%
\begin{eqnarray*}
T &=&\left( 
\begin{tabular}{cc}
$1$ & $0$ \\ 
$0$ & $\sqrt{\lambda _{k}}$%
\end{tabular}%
\right) ,\qquad \mathcal{R}_{\text{RG}}^{(11)}=1+\frac{\alpha _{k}}{2}(5-4%
\tilde{\gamma}_{M})-\frac{3\lambda _{k}}{2\varrho }, \\
\mathcal{R}_{\text{RG}}^{(22)} &=&1+\frac{\alpha _{k}}{12}(9-40\varrho )-%
\tilde{\gamma}_{M}\alpha _{k}(1-2\varrho )+\frac{3(1+2\varrho )\lambda _{k}}{%
4\varrho (1-2\varrho )}.
\end{eqnarray*}

Finally, the matrix $\mathcal{P}$ of perturbation spectra is defined from%
\begin{equation*}
\left\langle \mathcal{R}_{\text{RG}\mathbf{k}}(\tau )\hspace{0.01in}\mathcal{%
R}_{\text{RG}\mathbf{k}^{\prime }}^{\text{T}}(\tau )\right\rangle =(2\pi
)^{3}\delta ^{(3)}(\mathbf{k}+\mathbf{k}^{\prime })\frac{2\pi ^{2}}{k^{3}}%
\mathcal{P}.
\end{equation*}%
The result is%
\begin{equation}
\mathcal{P}=\frac{m_{\phi }^{2}G}{12\pi }\left( 
\begin{tabular}{cc}
$\mathcal{\tilde{P}}_{11}$ & $\mathcal{\tilde{P}}_{12}$ \\ 
$\mathcal{\tilde{P}}_{21}$ & $\mathcal{\tilde{P}}_{22}$%
\end{tabular}%
\right) ,  \label{calP}
\end{equation}%
where%
\begin{eqnarray}
\mathcal{\tilde{P}}_{11} &=&\frac{1}{\alpha _{k}^{2}}\left[ 1+\alpha
_{k}(5-4\gamma _{M})+\frac{3\lambda _{k}(1-2\varrho )}{2\varrho ^{2}}\right]
+\mathcal{O}_{0},\qquad \mathcal{\tilde{P}}_{12}=\mathcal{\tilde{P}}_{21}=%
\sqrt{\frac{3}{2}}\frac{1-\varrho }{\varrho \alpha _{k}}+\mathcal{O}_{0}, 
\notag \\
\mathcal{\tilde{P}}_{22} &=&\frac{1}{\lambda _{k}}\left\{ 1+(9-40\varrho )%
\frac{\alpha _{k}}{6}-2\alpha _{k}\gamma _{M}(1-2\varrho )+\frac{3}{2}%
\lambda _{k}\left( 1+\frac{1}{\varrho }+\frac{4}{1-2\varrho }\right)
\right\} +\mathcal{O}_{1}.  \label{pp}
\end{eqnarray}

Again, the spectra are manifestly RG invariant. In passing, the results
prove that $\alpha $ and $\varsigma =\sqrt{\alpha \lambda }$ are not the
right couplings, since if we used them, we would have negative powers of $%
\alpha _{k}$ in the expansion, and not just as overall factors. This shows
that it is not possible to identify the right couplings just from the
equations of the background metric and the background field $\varphi $.

We conclude by showing that the standard curvature perturbations $\mathcal{R}%
_{\text{adiab}}$ and $\mathcal{R}_{\text{iso}}$ of formulas (\ref{adiso})
are not the right physical quantities. They can be worked out from equations
(\ref{a1}), (\ref{a2}), then (\ref{psib}) (which express them in terms of $%
U_{1}$ and $U_{2}$), then (\ref{Uu}) (which relate $U_{1}$ and $U_{2}$ to $%
u_{1}$ and $u_{2}$), and finally (\ref{suso}) and (\ref{Qt0}) (which give $w=%
\sqrt{k}\left( u_{1}\text{, }u_{2}\right) $). We find, to the NLL order, 
\begin{eqnarray}
\mathcal{R}_{\text{adiab}} &=&\frac{im_{\phi }}{k^{3/2}\alpha _{k}}\sqrt{%
\frac{\pi G}{6}}\left( 1+\frac{\alpha _{k}}{2}(5-4\tilde{\gamma}_{M})-\frac{3%
}{2}\lambda _{k}+6\ell (1-\varrho )\lambda _{k}+\mathcal{O}_{2}\right. , 
\notag \\
&&\left. \sqrt{\frac{3\lambda _{k}}{2}}(1-4\ell (1-\varrho ))+\mathcal{O}%
_{3/2}\right) (\hat{A}_{\mathbf{k}}-\hat{A}_{-\mathbf{k}}^{\dag }),  \notag
\\
\mathcal{R}_{\text{iso}} &=&-\frac{im_{\phi }}{k^{3/2}\alpha _{k}}\sqrt{%
\frac{\pi G}{6}}(1+\ell +2\ell \varrho )\left( -\sqrt{\frac{3\lambda _{k}}{2}%
}+\mathcal{O}_{3/2},1+\mathcal{O}_{1}\right) (\hat{A}_{\mathbf{k}}-\hat{A}_{-%
\mathbf{k}}^{\dag }),  \label{rrr}
\end{eqnarray}%
up to corrections that contain higher powers $\alpha _{k}\ln \eta \equiv
\ell $. We recall that $\ell $ must be considered of order unity when we
make a log expansion, while it is of order one when we make an ordinary
expansion in powers of $\alpha _{k}$ and $\lambda _{k}$. The terms
proportional to $\ell $ of formulas (\ref{rrr}) are reported to show that $%
\ell $ does not disappear. This proves that $\mathcal{R}_{\text{adiab}}$ and 
$\mathcal{R}_{\text{iso}}$ are not RG invariant (i.e., they are not $\eta $
independent)\ in the superhorizon limit, differently from the combinations $%
\mathcal{R}_{\text{RG}}$ identified above. For this reason, they are not the
correct physical perturbations. The violation of RG invariance starts from
the subleading corrections in $\mathcal{R}_{\text{adiab}}$, but also affects
the leading contributions to $\mathcal{R}_{\text{iso}}$.

\section{Predictions}

\label{predictions}\setcounter{equation}{0}

In this section we comment on the predictivity of the double-field model.
Formula (\ref{llrc}) gives, to the leading order,%
\begin{equation}
\alpha _{k}\sim \frac{\alpha _{\ast }}{1+2\alpha _{\ast }\ln (k_{\ast }/k)}%
,\qquad \lambda _{k}\sim \frac{\bar{\lambda}}{\alpha _{\ast }^{1-2\varrho }}%
\alpha _{k}^{1-2\varrho },  \label{runna}
\end{equation}%
where $k_{\ast }$ is some pivot scale and $\alpha _{\ast }$ is the pivot
coupling. We see that when $\alpha _{k}$ is small, $\lambda _{k}$ is
generically larger, assuming that the constant $\bar{\lambda}/\alpha _{\ast
}^{1-2\varrho }$ is of order unity. Then, formulas (\ref{pp}) imply that the
leading contribution to the scalar spectrum is given by the 1-1 entry of the
matrix $\mathcal{P}$, while the rest can be treated as a correction. The
spectra to the leading order are%
\begin{equation*}
\mathcal{P}_{T}\simeq \frac{4m_{\phi }^{2}\zeta G}{\pi }(1-3\zeta \alpha
_{k}),\qquad \mathcal{P}_{11}\simeq \frac{m_{\phi }^{2}G}{12\pi \alpha
_{k}^{2}},
\end{equation*}%
while the tilts and the tensor-to-scalar ratio $r$ read%
\begin{equation}
n_{\text{t}}\simeq -6\zeta \alpha _{k}^{2},\qquad n_{\text{s}}-1\simeq
-4\alpha _{k},\qquad r=\frac{\mathcal{P}_{T}}{\mathcal{P}_{11}}\simeq \frac{%
96\alpha _{k}^{2}m_{\chi }^{2}}{2m_{\chi }^{2}+m_{\phi }^{2}}\simeq \frac{%
6(n_{\text{s}}-1)^{2}m_{\chi }^{2}}{2m_{\chi }^{2}+m_{\phi }^{2}}.
\label{predi}
\end{equation}%
Recalling that, by the ABP bound, $m_{\chi }$ must lie in the interval $%
m_{\phi }/4<m_{\chi }<\infty $, we conclude that $r$ lies the interval 
\begin{equation}
\frac{(n_{\text{s}}-1)^{2}}{3}\lesssim r\lesssim 3(n_{\text{s}}-1)^{2},
\label{preda}
\end{equation}%
at least in the perturbative regime that we have identified in this paper.
This result coincides with the one of pure quantum gravity found in \cite%
{ABP}. Moreover, in \cite{AFP} it was shown that the same prediction holds
when, in single-field inflation, the renormalizability requirement is
relaxed and the Starobinsky potential (which is equivalent to the $R^{2}$
term in (\ref{sqg})) is replaced by any potential of class I (as per the
classification of \cite{AFP}, which means a power series $\mathcal{V}(%
\mathrm{e}^{-c\phi })$ in $\mathrm{e}^{-c\phi }$, where $c$ is a constant, $%
\phi $ is the inflaton and $\mathcal{V}(0)\neq 0$). These observations
suggest that the prediction (\ref{preda}) is a robust prediction of quantum
gravity.

The second scalar affects the subleading corrections. If we define $n_{\text{%
s}}-1$ from $\mathcal{P}_{11}$, we find 
\begin{eqnarray*}
n_{\text{s}}-1 &=&-\beta _{\alpha }(\alpha _{k},\lambda _{k})\frac{\partial
\ln \mathcal{P}_{11}}{\partial \alpha _{k}}-\beta _{\lambda }(\alpha
_{k},\lambda _{k})\frac{\partial \ln \mathcal{P}_{11}}{\partial \lambda _{k}}
\\
&\simeq &-4\alpha _{k}+\frac{4}{3}\alpha _{k}^{2}(5-6\gamma _{M})+\frac{%
3(1-2\varrho )}{\varrho ^{2}}\alpha _{k}\lambda _{k}+\mathcal{O}_{3}.
\end{eqnarray*}%
Depending on the value of $\varrho =m^{2}/m_{\phi }^{2}$, the correction
proportional to $\alpha _{k}\lambda _{k}$ can be larger than the one
proportional to $\alpha _{k}^{2}$.

If the constant $\bar{\lambda}/\alpha _{\ast }^{1-2\varrho }$ is large
enough the analysis just made does not hold and it is possible to have a
dominant contribution from $\varphi $ to the scalar spetrum.

We point out that the conclusions just reached are nontrivial consequences
of RG invariance. To some extent, they are also unexpected. To illustrate
this point, let us assume for the moment that $\mathcal{R}_{\text{adiab}}$
and $\mathcal{R}_{\text{iso}}$ are the correct scalar perturbations, as one
would naively do. Then formulas (\ref{rrr}) show that to the leading order
they both behave like $1/\alpha _{k}$. However, $\mathcal{R}_{\text{iso}}$
has an extra factor $1+\ell +2\ell \varrho $. If we take $\mathcal{R}_{\text{%
iso}}$ in the tensor-to-scalar ratio $r$, we obtain%
\begin{equation}
\mathcal{P}_{T}\simeq \text{as above},\qquad \mathcal{P}_{\text{iso}}\simeq 
\frac{m_{\phi }^{2}G}{12\pi \alpha _{k}^{2}}(1+\ell +2\ell \varrho
)^{2},\qquad n_{\text{s}}-1=\frac{\mathrm{d}\ln \mathcal{P}_{\text{iso}}}{%
\mathrm{d}\ln k}\simeq -\frac{2(1-2\varrho )\alpha _{k}}{1+\ell +2\ell
\varrho },  \label{affa}
\end{equation}%
hence%
\begin{equation}
r\simeq \frac{48\zeta \alpha _{k}^{2}}{(1+\ell +2\ell \varrho )^{2}}\simeq 
\frac{24m_{\chi }^{2}(n_{\text{s}}-1)^{2}}{(1-2\varrho )^{2}(2m_{\chi
}^{2}+m_{\phi }^{2})}.  \label{affa2}
\end{equation}%
With respect to (\ref{predi}), $r$ is multiplied by an extra factor $%
4/(1-2\varrho )^{2}$, which can be as large as we wish if $\varrho $ is
close to its bound (\ref{bond}). We conclude that $\mathcal{R}_{\text{adiab}%
} $ and $\mathcal{R}_{\text{iso}}$ do not provide a meaningful prediction
for $r$ with the data available at present.

\section{Conclusions}

\label{conclusions}\setcounter{equation}{0}

We have shown that there are situations where we can import RG\ techniques
from high-energy physics into double-field inflation and study it as an
asymptotically de Sitter RG\ flow in two couplings. This means that there
exists a perturbative region where the cosmic RG flow resembles the one of
an asymptotically free quantum field theory.

The tensor perturbations are RG\ invariant in the superhorizon limit, but
the usual adiabatic and isocurvature perturbations are not, due to the
nontrivial mixing between the scalar fields. Nonetheless, RG invariance
allows us to work out the correct curvature perturbations algorithmically.

We worked out the power spectra of the tensor perturbations to the NNLL
order and the ones of the scalar perturbations to the NLL order. An
unexpected consequence of RG\ invariance is that, under mild assumptions,
the theory remains predictive and the predictions to the leading order
confirm those of pure quantum gravity. This does not occur if we use the
adiabatic and isocurvature perturbations as commonly defined. The results
suggest that, very much like gauge invariance, cosmic RG invariance is a
guiding principle to identify the right physical quantities.


Referring to the classification of potentials and cosmic RG flows done in
ref. \cite{AFP}, the model studied here describes a combination of a flow of
class I (due to the Starobinsky potential) with a flow of class II (due to
the powerlike $\varphi $ potential). The asymmetry between the two scalar
perturbations is due to the renormalizability requirement, because there is
no way to combine two flows of class I (or two flows of class II) into a
renormalizable theory. To further appreciate the features of double- and
multi-field inflation in simpler setups, it may be interesting to renounce
renormalizability and study symmetric configurations.

\vskip25truept \noindent {\large \textbf{Acknowledgments}}

\vskip 1truept

We thank D. Comelli, A. Karam and M. Piva for helpful discussions.

\vskip25truept \noindent {\LARGE \textbf{Appendices: reference formulas}}

\vskip20truept

\renewcommand{\thesection}{\Alph{section}} \renewcommand{\theequation}{%
\thesection.\arabic{equation}} \setcounter{section}{0}

In these appendices we collect involved formulas that are referred to in the
paper.

\vskip13truept

\section{Cosmic RG\ flow}

\label{formulas} \setcounter{equation}{0}

We start with some formulas used in the formulation of the cosmic RG flow of
section \ref{betaf}, which are the first contributions to the functions $v$, 
$\varphi $ and $H$ of expressions (\ref{vHph}):%
\begin{eqnarray}
v(\alpha ,\lambda ) &=&1-\alpha +\alpha ^{2}\left[ -2-\frac{29\alpha }{3}-%
\frac{638}{9}\alpha ^{4}+3\lambda \left( \frac{1}{\varrho }-2\right) \right. 
\notag \\
&&\left. +\alpha \lambda \left( \frac{30}{\varrho }-67+16\varrho \right) -%
\frac{9\lambda ^{2}}{\varrho }+\alpha ^{2}\mathcal{O}_{3}\right] ,  \notag \\
\frac{\hat{\kappa}\sqrt{\alpha }}{\sqrt{\lambda }}\varphi (\alpha ,\lambda )
&=&-\frac{1}{\varrho }+\frac{2\alpha }{3}-\frac{\alpha }{6\varrho }+\frac{%
3\lambda }{2\varrho ^{2}}-\frac{\alpha ^{2}}{3}\left( 2-\frac{4\varrho }{3}-%
\frac{5}{12\varrho }\right) +\frac{\alpha \lambda }{\varrho }-\frac{9\lambda
^{2}}{4\varrho ^{3}}  \notag \\
&&+\alpha ^{2}\lambda \left( \frac{8}{3}-\frac{11}{6\varrho }+\frac{1}{%
8\varrho ^{2}}\right) +\frac{3\alpha \lambda ^{2}}{8\varrho ^{3}}\left(
1-4\varrho \right)   \notag \\
&&+\frac{\alpha ^{3}}{216}\left( 304-\frac{57}{\varrho }-336\varrho
+128\varrho ^{2}\right) +\frac{27\lambda ^{3}}{8\varrho ^{4}}+\mathcal{O}%
_{4},  \label{vfiH} \\
\frac{\sqrt{6\alpha }}{m_{\phi }}H(\alpha ,\lambda ) &=&1-\frac{\text{$%
\alpha $}}{12}-\frac{3\lambda }{4\varrho }+\frac{19\text{$\alpha $}^{2}}{288}%
-\frac{\text{$\alpha \lambda $}}{2}\left( 1-\frac{3}{8\varrho }\right) +%
\frac{27\lambda ^{2}}{32\varrho ^{2}}-\frac{373\alpha ^{3}}{3456}  \notag \\
&&+\text{$\alpha $}^{2}\lambda \left( \frac{11}{8}-\frac{53}{128\varrho }%
-\varrho \right) +\frac{3\text{$\alpha $}\lambda ^{2}}{8\varrho }\left( 1-%
\frac{15}{16\varrho }\right) -\frac{135\lambda ^{3}}{128\varrho ^{3}}+%
\mathcal{O}_{4}.  \notag
\end{eqnarray}

\vskip13truept

\section{Tensor spectrum}

\setcounter{equation}{0}\label{apptens}

Here we collect formulas about the spectrum of the tensor perturbations
studied in section \ref{tensorQG}. We start with the coefficients of the
de-Sitter diagonalized Lagrangians of formulas (\ref{St}) and (\ref{diag}),
which are%
\begin{eqnarray}
&&q_{2}=\frac{1}{\tau ^{2}}-\frac{k^{2}}{2}+\left( \frac{9}{2\tau ^{2}}+\xi
k^{2}\right) \zeta \alpha ^{2}+\left[ \frac{3\left( 32+37\xi +6\xi
^{2}\right) }{\tau ^{2}}-\frac{k^{2}\xi }{3}(14-11\xi )\right] \frac{\zeta
^{2}}{4}\alpha ^{3}  \notag \\
&&\qquad +\left[ \frac{27}{2\tau ^{2}}+3k^{2}\xi \left( 2-\frac{1}{\varrho }%
\right) \right] \frac{\zeta }{2}\alpha ^{2}\lambda ,\qquad \qquad q_{4}=%
\frac{2}{\xi \tau ^{2}}+\frac{k^{2}}{2}+\frac{6\alpha }{\xi \tau ^{2}}, 
\notag
\end{eqnarray}%
\begin{eqnarray}
&&q_{5}=-\frac{\left( 2-37\xi +14\xi ^{2}\right) \zeta \alpha ^{2}}{4\xi
\tau ^{2}}+2k^{2}\xi \zeta \alpha ^{2}+\frac{9\alpha \lambda }{2\varrho \xi
\tau ^{2}}  \notag \\
&&-\frac{\left( 19-239\xi -182\xi ^{2}+42\xi ^{3}\right) \zeta ^{2}\alpha
^{3}}{12\xi \tau ^{2}}-\frac{1}{6}k^{2}\left( 14-11\xi \right) \xi \zeta
^{2}\alpha ^{3}  \label{appa} \\
&&+\frac{3\left( 2(3+\varrho )+2\left( 2\varrho +3\right) \xi +3\left(
1-2\varrho \right) \xi ^{2}\right) \zeta \alpha ^{2}\lambda }{2\varrho \xi
\tau ^{2}}+3k^{2}\xi \zeta \left( 2-\frac{1}{\varrho }\right) \alpha
^{2}\lambda ,  \notag \\
&&q_{6}=-\frac{\zeta \alpha ^{2}(1-\xi )}{3\tau }\left[ 6+5\alpha +9\left( 2-%
\frac{1}{\varrho }\right) \lambda \right] .  \notag
\end{eqnarray}

\paragraph{Fakeon projection.}

The coefficients of the fakeon projection (\ref{VU}) read%
\begin{eqnarray}
s_{1}(\alpha ,\lambda ) &=&\frac{\zeta \alpha }{16}\left[ \zeta \alpha
\left( 2-29\xi +6\xi ^{2}\right) -\frac{18\lambda }{\varrho }\right] +\frac{%
\zeta ^{3}\alpha ^{3}}{48}(1+132\xi -303\xi ^{2}+62\xi ^{3})  \notag \\
&&+\frac{3\zeta ^{2}\alpha ^{2}\lambda }{8\varrho }(3-5\xi -\xi ^{2}-\varrho
(2+3\xi -2\xi ^{2}))+\alpha \mathcal{O}_{3},  \label{s12} \\
s_{2}(\alpha ,\lambda ) &=&-\frac{\xi (1-\xi )}{2}\zeta ^{2}\alpha ^{2}+%
\frac{\xi \zeta ^{3}\alpha ^{3}}{24}(32-130\xi +35\xi ^{2})-\frac{3\xi \zeta
^{2}\alpha ^{2}\lambda }{4\varrho }(2+\xi -\varrho (1+2\xi ))+\alpha 
\mathcal{O}_{3}.  \notag
\end{eqnarray}

\paragraph{RG invariance.}

The solution of equation (\ref{DQt}) is (\ref{Qtildet}) with 
\begin{eqnarray}
J_{\text{t}}(\alpha ,\lambda ) &=&1+\frac{3\zeta \alpha }{2}+\frac{56+61\xi
+12\xi ^{2}}{16}\zeta ^{2}\alpha ^{2}+\frac{9}{8\varrho }\zeta \alpha \lambda
\notag \\
&&+\left( 1544+2147\xi +1052\xi ^{2}+162\xi ^{3}\right) \frac{\zeta
^{3}\alpha ^{3}}{96}  \label{Jt} \\
&&-\frac{3}{16\varrho }\left[ 13+14\xi +6\xi ^{2}-6\varrho (6+7\xi +2\xi
^{2})\right] \zeta ^{2}\alpha ^{2}\lambda +\mathcal{O}_{4}.  \notag
\end{eqnarray}%
The constants $c_{i}$ that parametrize the $\mathcal{O}_{4}$ contributions (%
\ref{o4sigmat}) to $\sigma _{\text{t}}$ read 
\begin{eqnarray}
c_{1} &=&277+\frac{3109}{8}\xi +\frac{345}{4}\xi ^{2}+81\zeta ,\qquad c_{3}=%
\frac{81}{4\zeta \varrho },  \notag \\
c_{2} &=&-\frac{9}{8\varrho }\left[ 112+107\xi +30\xi ^{2}-\varrho
(262+293\xi +72\xi ^{2})+8\varrho ^{2}(8+10\xi +3\xi ^{2})\right] .
\label{c3}
\end{eqnarray}

\paragraph{Normalization of the spectrum.}

The constant $\tilde{Q}_{\text{t}}(\alpha _{k},\lambda _{k})$ of (\ref%
{Qtildet}) is%
\begin{eqnarray}
\tilde{Q}_{\text{t}}(\alpha _{k},\lambda _{k}) &=&\frac{i}{\sqrt{2}}\left[ 1+%
\frac{3\zeta \alpha _{k}^{2}}{2}(4+\xi -2\tilde{\gamma}_{M})\right. -\zeta
\alpha _{k}^{3}\pi ^{2}-6\zeta \alpha _{k}^{3}\tilde{\gamma}_{M}^{2}+\frac{3%
}{2}\tilde{\gamma}_{M}\zeta ^{2}\alpha _{k}^{3}(8+\xi )  \notag \\
&&\left. +\frac{9}{8}\xi \zeta ^{2}\alpha _{k}^{3}(10+3\xi )+\frac{9}{4}%
\zeta \alpha _{k}^{2}\lambda _{k}\left( 4-2\tilde{\gamma}_{M}+2\xi -\frac{%
\xi }{\varrho }\right) +\mathcal{O}_{4}\right] .  \label{Qwt}
\end{eqnarray}

\section{Scalar spectrum}

\setcounter{equation}{0}\label{appscal}

The solutions of the MS equations derived from (\ref{SMSt}) with the
expansion (\ref{Utilde}) are given by%
\begin{eqnarray}
P_{0} &=&W_{0}\mathbb{I},\quad P_{1}=\frac{W_{2}}{3}\left( \!%
\begin{tabular}{cc}
$2$ & $0$ \\ 
$0$ & $1-2\varrho $%
\end{tabular}%
\!\right) ,\quad P_{2}=\sqrt{\frac{3}{2}}W_{2}\left( \!%
\begin{tabular}{ll}
$0$ & $1$ \\ 
$1$ & $0$%
\end{tabular}%
\!\right) ,\quad P_{4}=-\frac{W_{2}}{\varrho }\left( \!%
\begin{tabular}{ll}
$1$ & $0$ \\ 
$0$ & $\varrho $%
\end{tabular}%
\!\right) ,  \notag \\
P_{3} &=&\frac{W_{4}}{4}\left( 
\begin{tabular}{cc}
$4$ & $0$ \\ 
$0$ & $(1-2\varrho )^{2}$%
\end{tabular}%
\ \right) +\frac{1}{6}\left[ \frac{1}{2}(1-4\varrho ^{2})W_{3}+\frac{%
5+18\varrho -12\varrho ^{2}}{3}W_{2}\right] \left( 
\begin{tabular}{ll}
$0$ & $0$ \\ 
$0$ & $1$%
\end{tabular}%
\right) .  \label{scalauto}
\end{eqnarray}

\section{Adiabatic and isocurvature perturbations}

\setcounter{equation}{0}\label{adiabiso}

The adiabatic and isocurvature perturbations can be defined by switching to
the inflaton framework through a Weyl transformation combined with a
diffeomorphism (see the appendix of \cite{ABP} for details). The latter
amounts to defining a new cosmological time $\bar{t}$, to preserve the
structure of the parametrization (\ref{metr}) of the metric. We have 
\begin{equation}
g_{\mu \nu }\mathrm{d}x^{\mu }\mathrm{d}x^{\nu }=\mathrm{e}^{2\sigma }\bar{g}%
_{\mu \nu }\mathrm{d}\bar{x}^{\mu }\mathrm{d}\bar{x}^{\nu },\qquad \mathrm{e}%
^{-2\sigma }=1-\frac{\Omega _{0}+\Omega }{3m_{\phi }^{2}},\qquad \frac{\text{%
\textrm{d}}\bar{t}}{\text{\textrm{d}}t}=\mathrm{e}^{-\sigma _{0}},
\label{a1}
\end{equation}%
where the bars are used to denote quantities in the inflaton framework. The
space coordinates are not affected by the transformation.

From the transformed action (\ref{acc}), we find the scalar fields $\phi
^{i}=(\sigma ,\varphi )=(\sigma _{0}+\delta \sigma ,\Theta _{0}+\Theta
)\equiv \phi _{0}^{i}+\delta \phi ^{i}$ and\ the matrix $G_{ij}(\phi _{0})=$
diag$(4/\hat{\kappa}^{2},\mathrm{e}^{2\sigma _{0}})$, defined so that the
scalar kinetic terms of the new action read%
\begin{equation*}
\frac{1}{2}\int \mathrm{d}^{4}\bar{x}\sqrt{-\bar{g}}G_{ij}(\phi )\bar{g}%
^{\mu \nu }\bar{\partial}_{\mu }\phi ^{i}\bar{\partial}_{\nu }\phi ^{j}.
\end{equation*}%
The formulas of \cite{ABP} give 
\begin{equation}
\bar{\Psi}=\delta \sigma =\frac{\Omega }{6m_{\phi }^{2}}\mathrm{e}^{2\sigma
_{0}}.  \label{a2}
\end{equation}

The standard definitions of adiabatic and isocurvature perturbations are 
\cite{gundhi}%
\begin{equation}
\mathcal{R}_{\text{adiab}}=\bar{\Psi}+\bar{H}\frac{G_{ij}(\phi
_{0})(\partial _{\bar{t}}\phi _{0}^{i})\delta \phi ^{j}}{G_{kl}(\phi
_{0})(\partial _{\bar{t}}\phi _{0}^{k})(\partial _{\bar{t}}\phi _{0}^{l})}%
,\qquad \mathcal{R}_{\text{iso}}=\frac{G_{ij}(\phi _{0})\psi _{0}^{i}(\delta
\phi ^{j}\bar{H}+\bar{\Psi}\partial _{\bar{t}}\phi _{0}^{j})}{\sqrt{%
G_{kl}(\phi _{0})(\partial _{\bar{t}}\phi _{0}^{k})(\partial _{\bar{t}}\phi
_{0}^{l})}},  \label{adiso}
\end{equation}%
where $\psi _{0}^{i}$ are defined so that $G_{ij}(\phi _{0})\psi
_{0}^{i}\psi _{0}^{j}=1$ and $G_{ij}(\phi _{0})\psi _{0}^{i}(\partial _{\bar{%
t}}\phi _{0}^{j})=0$.

The explicit expressions of $\mathcal{R}_{\text{adiab}}$ and $\mathcal{R}_{%
\text{iso}}$, which we do not report here, can be worked out from the
formulas just given. We have used them to work out (\ref{rrr}) and the
predictions (\ref{affa}) and (\ref{affa2}) of section \ref{predictions}.

\section{Other formulas}

\setcounter{equation}{0}\label{otherf}

The functions%
\begin{eqnarray}
W_{0} &=&\frac{i(1-i\eta )}{\eta \sqrt{2}}\mathrm{e}^{i\eta },\qquad W_{2}=%
\frac{6W_{0}}{1-i\eta }-3\left( i\pi -\hspace{0.01in}\text{Ei}(2i\eta
)\right) W_{0}^{\ast },  \notag \\
W_{3} &=&\left[ 6(\ln \eta +\tilde{\gamma}_{M})^{2}+24i\eta
F_{2,2,2}^{1,1,1}\left( 2i\eta \right) +\pi ^{2}\right] W_{0}^{\ast }+\frac{%
24W_{0}}{1-i\eta }-4(\ln \eta +1)W_{2},  \label{stada} \\
W_{4} &=&-\frac{16W_{0}}{1+\eta ^{2}}+\frac{2(13+i\eta )W_{2}}{9(1+i\eta )}+%
\frac{W_{3}}{3}+4G_{2,3}^{3,1}\left( -2i\eta \left\vert _{0,0,0}^{\
0,1}\right. \right) W_{0},  \notag
\end{eqnarray}%
introduced in \cite{FakeOnScalar} appear frequently and can be used to
express the solutions of the Mukhanov-Sasaki equations to the lowest orders.
Ei denotes the exponential-integral function, $F_{b_{1},\cdots
,b_{q}}^{a_{1},\cdots ,a_{p}}(z)$ the generalized hypergeometric function $%
_{p}F_{q}(\{a_{1},\cdots ,a_{p}\},\{b_{1},\cdots ,b_{q}\};z)$ and $%
G_{p,q}^{m,n}$ the Meijer G function.

The asymptotic behaviors in the superhorizon limit $\eta \simeq 0$ are%
\begin{eqnarray}
\eta W_{2} &\simeq &\frac{3i}{\sqrt{2}}\left( 2-\tilde{\gamma}_{M}-\ln \eta
\right) ,\qquad \eta W_{4}\simeq -8i\sqrt{2}+\frac{26}{9}\eta W_{2}+\frac{%
\eta W_{3}}{3}+i\sqrt{2}(\ln \eta +\tilde{\gamma}_{M})^{2}+\frac{i\pi ^{2}}{%
\sqrt{2}},  \notag \\
\eta W_{3} &\simeq &-3i\sqrt{2}(\ln \eta +\tilde{\gamma}_{M})^{2}-\frac{i\pi
^{2}}{\sqrt{2}}+12i\sqrt{2}-4(\ln \eta +1)\eta W_{2}.  \label{beha}
\end{eqnarray}

We quantize (\ref{mu}) as usual, by introducing the operator 
\begin{equation*}
\hat{w}_{\mathbf{k}}(\eta )=w_{\mathbf{k}}(\eta )\hat{a}_{\mathbf{k}}+w_{-%
\mathbf{k}}^{\ast }(\eta )\hat{a}_{-\mathbf{k}}^{\dagger },
\end{equation*}%
where $\hat{a}_{\mathbf{k}}^{\dagger }$ and $\hat{a}_{\mathbf{k}}$ are
creation and annihilation operators satisfying $[\hat{a}_{\mathbf{k}},\hat{a}%
_{\mathbf{k}^{\prime }}^{\dagger }]=(2\pi )^{3}\delta ^{(3)}(\mathbf{k}-%
\mathbf{k}^{\prime })$, $[\hat{a}_{\mathbf{k}},\hat{a}_{\mathbf{k}^{\prime
}}]=[\hat{a}_{\mathbf{k}}^{\dag },\hat{a}_{\mathbf{k}^{\prime }}^{\dagger
}]=0$, so that%
\begin{equation*}
\langle w_{\mathbf{k}}(\eta )w_{\mathbf{k}^{\prime }}(\eta )\rangle =(2\pi
)^{3}\delta ^{(3)}(\mathbf{k}+\mathbf{k}^{\prime })|w_{\mathbf{k}}|^{2}.
\end{equation*}%
Summing over the graviton polarizations $u$ and $v$, the power spectrum $%
\mathcal{P}_{T}$ of the tensor perturbations is defined by the two-point
function%
\begin{equation}
\langle \hat{u}_{\mathbf{k}}(\tau )\hat{u}_{\mathbf{k}^{\prime }}(\tau
)\rangle =(2\pi )^{3}\delta ^{(3)}(\mathbf{k}+\mathbf{k}^{\prime })\frac{\pi
^{2}}{8k^{3}}\mathcal{P}_{T},\qquad \mathcal{P}_{T}=\frac{8k^{3}}{\pi ^{2}}%
|u_{\mathbf{k}}|^{2},  \label{ppt}
\end{equation}

For convenience, we recall some notations frequently used in the paper,
which are%
\begin{equation*}
\varrho =\frac{m^{2}}{m_{\phi }^{2}},\qquad \xi =\frac{m_{\phi }^{2}}{%
m_{\chi }^{2}},\qquad \zeta =\left( 1+\frac{\xi }{2}\right) ^{-1},\qquad 
\tilde{\gamma}_{M}=\gamma _{M}-\frac{i\pi }{2},\qquad \gamma _{M}=\gamma
_{E}+\ln 2,
\end{equation*}%
$\gamma _{E}$ being the Euler-Mascheroni constant.

\end{document}